\documentclass[accepted]{uai2026}
\input{Definitions}
                        
\usepackage[american]{babel}

\usepackage{natbib} 
\usepackage{mathtools} 
\usepackage{booktabs} 
\usepackage{tikz} 
\usepackage{adjustbox}
\usepackage{microtype}
\usepackage{graphicx}
\usepackage{pdfpages}
\usepackage{subcaption}
\usepackage{circledsteps}
\usepackage[ruled,vlined]{algorithm2e}

\title{Trading Utility for Dynamic Fairness in Multiple Resource Division with Sequential Demand}

\author[1]{\href{mailto:<kjiang10@uic.edu>?Subject=UAI 2026 paper}{Kaiqi~Jiang}{}}
\author[1]{Karim~El~Husseini}
\author[1]{Wenzhe~Fan}
\author[1]{Xinhua~Zhang}
\affil[1]{%
    Computer Science Dept.\\
    University of Illinois Chicago\\
    Chicago, Illinois, USA
}
  
\begin{document}
\maketitle
\begin{abstract}
Dynamic multi-resource allocation is a central problem in shared computing environments, where users' demands arrive sequentially and resources must be distributed fairly without knowledge of future demands. Existing methods emphasize fairness guarantees such as Sharing Incentive, Envy Freeness, and Dynamic Pareto Optimality, but often overlook system utility. Moreover, these fairness criteria are mutually incompatible, preventing strict enforcement of them at the same time. \\

We propose a neural allocation mechanism that reconciles fairness with utility through multi-objective optimization during sequential rollout. We first formalize fairness in the dynamic setting via stepwise loss functions for Sharing Incentive, Envy Freeness, and Dynamic Pareto Optimality, enabling differentiable training. Leveraging non-wastefulness, we parameterized the solutions by constraining allocations to the subspace of demand while allowing elastic over-allocation when resources remain available.  Empirical results demonstrate that our learned allocator achieves substantially higher utility at comparable levels of fairness,
uncovering clear Pareto-frontier-like tradeoffs across metrics. 
\end{abstract}

\vspace{-0.75em}
\section{Introduction}

Dynamic multi-resource allocation is a fundamental problem in shared computing platforms such as cloud clusters and GPU pools, where users arrive sequentially and request heterogeneous resources such as CPUs, GPUs, memory, etc. The allocator must make decisions without knowledge of future demands while ensuring fairness across users and time.
This is in contrast to the static setting where all users' demands are presented at the same time. Classical works in fair division have established axiomatic criteria such as \textit{Sharing Incentive (SI)}, \textit{Envy Freeness (EF)}, and \textit{Dynamic Pareto Optimality (DPO)} as desirable fairness notions in multi-agent allocation \citep{moulin2003fair, walsh2011online}. However, in general, these criteria are mutually incompatible in dynamic settings, making it impossible to enforce all of them simultaneously \citep{kash2013no}.

A central milestone in multi-resource allocation was the introduction of \textit{Dominant Resource Fairness (DRF)} by \citet{ghodsi2011dominant}, which equalizes users’ dominant shares across multiple resource types in static environments. Since then, numerous extensions have been developed, including generalized utility models \citep{gutman2012fair}, and queue-based systems \citep{ghodsi2012multiresource}. These approaches provide strong fairness guarantees but give limited attention to \textbf{utility}, i.e., the degree to which resource allocations satisfy user demands and maximize system efficiency (we use utility and efficiency interchangeably). In both static and dynamic computing systems, utility is equally important: wasted or underutilized resources directly degrade overall throughput. In the \textit{static} setting, all users are present simultaneously and resources are allocated in one shot. In the \textit{dynamic} setting, however, users arrive sequentially, and the allocator must decide without knowledge of future demands. This sequential nature makes it substantially more challenging to balance fairness and utility.

In a slightly different setting, 
\citet{zeng2020fairness} characterized how much efficiency needs be sacrificed to achieve stronger fairness guarantees when \textbf{items} (resources) arrive sequentially,
while users employ a time-varying valuation function of items.
Their work focused on \textit{indivisible} resource items effectively of a \textit{single} type (e.g., CPU), 
providing worst-case trade-off bounds under envy-based notions of fairness. 
In contrast, we study \textit{divisible} \textit{multi}-resource allocation with sequential arrivals of \textbf{demand},
and we focus on \textit{empirical} insights into the achievable trade-offs among multiple fairness metrics and efficiency.


The literature on \textit{dynamic fair division} was initiated by \citet{kash2013no}, who established the impossibility results for jointly satisfying EF and DPO.
They hence proposed relaxed conditions such as Dynamic DRF and Cautious LP based on CDPO (Cautious DPO). 
This enabled simultaneous satisfaction,
and algorithms are designed to achieve it.
Similarly, \citet{walsh2011online} proposed new relaxations of fairness criteria for online cake cutting,
and showed their algorithmic feasibility.
However, a natural question arises: 
\textbf{instead of relaxing fairness conditions and satisfying them strictly,
can we retain the original conditions but achieve them only approximately}?

\vspace{-0.75em}
\paragraph{Our perspective}  Inspired by this intuition,
we part from conventional mechanisms and propose a \textit{neural allocation mechanism} that learns to balance fairness and efficiency dynamically. Rather than enforcing fairness notions as strict constraints, we cast them as multi-objective \textit{stepwise differentiable loss functions}, including SI, EF, and DPO. 
They are optimized along with utility.

Bringing utility into the learning objective, however, incurs extra computational cost.
A new scalar hyperparameter needs to be tuned that corresponds to utility,
and this is on top of the existing hyperparameters corresponding to multiple fairness losses.
\textbf{Our first contribution} is to discover, empirically, 
that certain fairness criteria are well correlated with utility, 
hence eligible to serve as a proxy.
In other words, by incorporating this fairness into the overall objective,
we can promote the utility for free.
EF does not qualify because it only concerns users' relative utilities instead of their absolute values.
SI simply compares with a weak alternative (uniform allocation), hence also not qualified.
In contrast, DPO asserts there is no Pareto-improving allocation that uses at most $k/N$ of every resource at step $k$ of a horizon $N$.
This aligns well with utility, and our experiments show that it serves as an effective proxy of utility.
Existing works have studied correlations between (static) fairness metrics, 
based on which representative metrics can be identified \citep{anahideh2025expert}.
However, little is studied for dynamic fairness.


\textbf{Secondly}, we simplify our allocation policy by enforcing non-wastefulness such that  allocations are proportional to demands.
With carefully selected input features,
our allocator is endowed with expressive power to navigate fairness–utility trade-offs beyond other baselines such as DRF. 
By conditioning on leftover resources. 
controlled over-allocation is permitted when resources are available.

Our experiments introduce an evaluation framework based on Pareto set that jointly characterizes fairness metrics and utility. 
Our approach highlights the multi-dimensional nature of fairness, 
enabling principled and interpretable comparisons between learned and classical handcrafted allocation algorithms.
%
Overall, our neural policy advances fairness-aware allocation beyond these handcrafted algorithms in practice, 
achieving strong fairness metrics while delivering significantly better efficiency in dynamic multi-resource environments.

\section{Related Work}

Fair division has a long history in economics and computer science, with foundational work by \citet{moulin2003fair},
along with \citet{walsh2011online} on online cake cutting. In the context of multi-resource allocation, the most widely adopted baseline is \emph{Dominant Resource Fairness (DRF)} \citep{ghodsi2011dominant}, which equalizes dominant shares and guarantees Pareto efficiency and DPO in static settings. DRF has also been extended to networking via multi-resource fair queuing \citep{ghodsi2012multiresource}. While influential, DRF’s guarantees degrade under sequential arrivals, and our experiments confirm that it often sacrifices utility to achieve low fairness losses.

This limitation led to a central theoretical advance: the impossibility result by \citet{kash2013no}, showing that no mechanism can simultaneously satisfy SI, EF, and DPO in dynamic environments, in particular EF versus DPO. This impossibility sparked subsequent work on relaxations such as CDPO, as well as studies on proportionality and envy freeness in indivisible settings \citep{amanatidis2022fair}, and temporal fairness in online decision-making \citep{torres2024temporal}. Together, these works clarified that fairness criteria inevitably conflict, but most were analyzed at a single fairness operating point without systematically mapping the fairness–efficiency space.

Building on these theoretical insights, later contributions began to examine the nature of trade-offs more directly. For example, \citet{friedman2015dynamic} introduced bounded disruptions, linking fairness guarantees to allowable reallocations, while \citet{benade2022dynamic} showed that partial information can still yield fairness with high probability. These results emphasize the need for flexible mechanisms that adjust fairness and efficiency dynamically rather than strictly enforcing one fairness notion.

A related line of work incorporates fairness by adopting fixed criteria in advance. For instance, Themis balances short-term efficiency with long-run finish-time fairness \citep{mahajan2020themis}, 
and Karma allocates credits to users under heterogeneous workloads \citep{vuppalapati2023karma}. Although effective in their contexts, such approaches remain committed to predetermined fairness definitions and do not explore the broader fairness–utility trade-off frontier.

Another strand of research has applied reinforcement learning to scheduling in cluster systems. While RL can improve throughput and job completion time, it often introduces instability and lacks principled fairness guarantees. In contrast, our approach entirely avoids RL: by formulating fairness losses in a differentiable way, we train with standard supervised optimization, which yields stable learning, higher utility, and competitive fairness guarantees.

Overall, prior works have either enforced fairness strictly, studied one fairness metric in isolation, or optimized for efficiency while leaving fairness implicit. Our contribution is to go beyond these limitations: by treating fairness notions as differentiable objectives, we directly expose the full Pareto set across SI, EF, DPO, and utility, providing a principled characterization of achievable trade-offs rather than a single fairness–efficiency balance.

\section{Preliminary}

We study the problem of \emph{dynamic fair division} in a multi-resource setting. A system has $m$ types of resources (e.g., CPU, memory, GPUs), each normalized to total capacity 1. There are $N$ users arriving sequentially, and upon arrival, each user $i \in [N] := \{1, \dots, N\}$ reports a demand vector $d_i \in \mathbb{R}^m_{+}$ normalized so that $\max_{r \in [m]} d_{ir} = 1$. An allocation mechanism must assign resources irrevocably at each step without knowledge of future arrivals.

\vspace{-0.5em}
\paragraph{Allocations}  
At step $k \in [N]$, the mechanism outputs a partial allocation $A^k \in \mathbb{R}^{k \times m}$ for the first $k$ users. The irrevocability condition requires that allocations are monotone across steps:
\begin{equation}
\label{eq:obj_monotone}
A^k_{ir} \;\ge\; A^{k-1}_{ir}, \qquad \forall\, i \le k-1, \; \forall\, r.
\end{equation}
When no replenishment is allowed, we have
$A^k_{ir} = A^{i}_{ir}$ for all $k \ge i$.

\vspace{-0.5em}
\subsection{Utility}
The utility of user $i \le k$ is defined by the minimum fraction of demand satisfied across all resources:
\begin{equation}
u_i(A^k_{i:}) := \min_{r} \frac{A^k_{ir}}{d_{ir}}.
\label{eq:utility}
\end{equation}
This ``dominant-share'' utility reflects the bottleneck resource for the user: utility is high only if all resources are satisfied proportionally.

\subsection{Fairness Criteria}
\label{subsec:fairness_criteria}
Several well-established axioms govern fairness in dynamic multi-resource division:

\vspace{-0.5em}
\paragraph{Sharing Incentives (SI)}  
Each user should prefer its allocation to an equal split of the system:
\begin{equation}
u_i(A^k_{i:}) \;\ge\; u_i\!\left(\tfrac{1}{N},\ldots,\tfrac{1}{N}\right), \qquad \forall\, i \le k.
\label{eq:si}
\end{equation}

\vspace{-0.5em}
\paragraph{Envy Freeness (EF)}  
No user should envy another user’s allocation:
\begin{equation}
u_i(A^k_{i:}) \;\ge\; u_i(A^k_{j:}), \qquad \forall\, i,j \le k.
\label{eq:ef}
\end{equation}

\vspace{-0.5em}
\paragraph{Dynamic Pareto Optimality (DPO)}  
At each step, there must exist at least one resource whose cumulative allocation scales with the number of served users:
\begin{equation}
\exists\, r \;\; \text{s.t.} \;\; \sum_{i=1}^{k} A^k_{ir} \;\ge\; \frac{k}{N}.
\label{eq:dpo}
\end{equation}

\subsection{Tension Among Fairness Notions}
It is well known that the above criteria cannot be satisfied simultaneously in dynamic settings. 
For instance, mechanisms that guarantee SI and EF may violate DPO, while those that achieve DPO may fail EF. 
This tension is captured by the following impossibility result for dynamic fair division:

\begin{theorem}[\citet{kash2013no}]
No dynamic allocation mechanism can simultaneously satisfy Envy Freeness and Dynamic Pareto Optimality.
\end{theorem}

This impossibility theorem underscores a fundamental limitation of rule-based approaches: one must inevitably trade off among fairness desiderata. 
In particular, enforcing one criterion strictly comes at the cost of violating another. 
This motivates frameworks that balance utility and fairness trade-offs, rather than attempting strict enforcement of all criteria.

\section{Methodology}

A key methodological insight in our framework is that we can reduce what would traditionally be modeled as a reinforcement learning (RL) problem into a supervised learning problem. The task of dynamic multi-resource allocation can be naturally framed as sequential decision making: at each step $k$, the allocation $A^k$ must be chosen given all previous arrivals and allocations. Therefore, \emph{the current action space depends on the previous actions}. A naive approach would treat this as an RL setting with delayed rewards and long-term dependencies. However, we show that RL is unnecessary: the problem can be reformulated as supervised optimization with differentiable objectives.

Formally, we aim to learn a deterministic policy $f$ that maps past information to the current allocation:
\begin{equation}
A^k \leftarrow f(N, d_{\leq k}, A^{k-1}).
\label{eq:policy}
\end{equation}

In our design, $f$ is instantiated as a neural allocation mechanism parameterized by a multi-layer perceptron. Its inputs are the step ratio, the current demand vector $d_k$, the leftover resources, and the previous allocation. Its output is a proportional allocation vector $A^k$ that respects feasibility by construction.

The reward function $r$ is defined as a surrogate objective that measures fairness--utility trade-offs:
\begin{equation}
r(N, d_{\leq k}, A^k),
\label{eq:reward}
\end{equation}
where $r$ is not a single metric but a weighted combination of differentiable fairness losses, namely SI, EF, and DPO. We defer the modified SI, EF, and DPO losses in subsection~\ref{subsec:fairness_losses}.

Let each demand $d_i$ be drawn i.i.d.\ from a distribution $p_i$. The overall training objective is then
\begin{equation}
\max_f \ \expunder{d_1 \sim p_1, \ldots, d_N \sim p_N} 
\sum_{k=1}^N r(N, d_{\leq k}, A^k),
\label{eq:objective}
\end{equation}
subject to
\begin{equation}
\label{eq:constraint}
A^k = f(k, N, d_{\leq k}, A^{k-1}).
\end{equation}
%
%
\textbf{Here, $f$ implicitly ensures some basic constraints are satisfied},
such as nonnegativity, monotonicity as in \eqref{eq:obj_monotone},  and no exceeding of available remaining resources.
In practice, we approximate this expectation by independently sampling $B$ sequences of demands, denoted $\{ d_k^{(b)} \}_{k=1}^N$ for $b \in [B]$. The empirical training objective becomes
\begin{equation}
\max_f \ \frac{1}{BN} \sum_{b=1}^B \sum_{k=1}^N 
r(N, d_{\leq k}^{(b)}, A^{(b),k}),
\label{eq:empirical}
\end{equation}
with constraints
\begin{align}
& A^{(b),k} = f(N, d_{\leq k}^{(b)}, A^{(b),k-1}), \ \forall b \in [B], \ k \in [N].
\label{eq:empirical-constraint}
\end{align}
This formulation reveals that the allocation problem does not require reinforcement learning. Instead, it suffices to sample sequences of demands, roll out allocations under the deterministic policy $f$, and compute stepwise fairness losses. These losses are differentiable and can be directly optimized via backpropagation. Long-range dependencies across steps (e.g., $A^N$ depending on $A^1$) are naturally handled by the sequential rollout.

In summary, we cast dynamic multi-resource allocation as supervised learning: the allocation function $f$ is implemented as a neural network, fairness principles are encoded into the reward $r$, and training proceeds via backpropagation on sampled sequences. This design is both expressive and stable, producing allocations that empirically balance fairness with significantly improved utility.

\subsection{Fairness Losses}
\label{subsec:fairness_losses}

We translate classical fairness notions introduced in subsection~\ref{subsec:fairness_criteria} into \textit{stepwise differentiable loss functions}, aggregated in a rollout fashion. This enables end-to-end training of allocation policies using standard optimization.

\paragraph{Utility}  
For a user $i$ with demand vector $d_i \in \mathbb{R}^m$ and allocation vector $A_i^k \in \mathbb{R}^m$, utility is defined as
\begin{equation}
    u_i^k = \min_{r \in [m]} \frac{A_{i,r}^k}{d_{i,r} + \epsilon},
    \label{eq:utility}
\end{equation}
where $\epsilon > 0$ prevents division by zero. The total utility reported in experiments is the average across all users and steps:
\begin{equation}
    \mathcal{U} = \frac{1}{N}\sum_{k=1}^N \rbr{\frac{1}{k} \sum_{i=1}^k u_i^k}.
    \label{eq:utility-total}
\end{equation}

\paragraph{Sharing Incentive (SI)}  
At step $k$, each user should obtain at least as much utility as under equal allocation. The SI loss is
\begin{equation}
    \ell_{\text{SI}}^k = \frac{1}{k} \sum_{i=1}^k 
    [u_i \left(\tfrac{1}{N},\ldots,\tfrac{1}{N}\right) - u_i^k]_+.
    \label{eq:si-loss}
\end{equation}
Here $[\cdot]_+ := \max(0,\cdot)$ is the ReLU activation.

\paragraph{Envy Freeness (EF)}  
At step $k$, no user $i$ should prefer the allocation of another user $j$. The EF loss is
\begin{equation}
    \ell_{\text{EF}}^k = \frac{1}{k^2} \sum_{i=1}^k \sum_{j=1}^k
    [u_i(A_j^k) - u_i(A_i^k)]_+.
    \label{eq:ef-loss}
\end{equation}

\paragraph{Dynamic Pareto Optimality (DPO)}  
At step $k$, Pareto optimality requires at least one resource to be saturated at the proportional level $k/N$. The DPO loss is
\begin{equation}
    \ell_{\text{DPO}}^k = 
    \sbr{\tfrac{k}{N} - \max_{r \in [m]} \sum_{i=1}^k A_{i,r}^k}_+.
    \label{eq:dpo-loss}
\end{equation}
 
\paragraph{Total Fairness Loss}  
The aggregated fairness loss is
\begin{align}
\label{eq:fair-loss}
    \mathcal{L}_{\text{fair}} 
    = \frac{1}{N} \sum_{k=1}^N \Big(
        \lambda_{\text{SI}} \,\ell_{\text{SI}}^k
        + \lambda_{\text{EF}} \,\ell_{\text{EF}}^k
        + \lambda_{\text{DPO}} \,\ell_{\text{DPO}}^k
    \Big).\!    
\end{align}
It is straightforward to further include the negative of utility $\mathcal{U}$,
which carries its own weight $\lambda_{\text{UT}}$.
We will present its results in the supplementary material.

\begin{algorithm}[t]
\caption{FairUtil allocator}
\label{alg:fair_util}
\KwIn{demands $\{d_k\}_{k \in [N]}$, total capacity $c_1 \in \RR^m \!\!\!\!\!$}
\KwOut{allocations $\{A^1, \dots, A^N\}$}

\For{$k = 1$ \KwTo $N$}{
    $r_k \gets \tfrac{k}{N}$\;
    $x_k \gets [\, r_k ; d_k ; c_k ; A^{k-1} \,]$\;
    $s_k \gets \text{Softplus}(g(x_k))$\;
    $\tilde{a}^k \gets s_k \cdot d_k$\;
    $a^k \gets \min(\tilde{a}^k, c_k)$, and $A^k = [a^1, \ldots, a^k]^\top$ \; 
    $c_{k+1} \gets c_k - a^k$\;
}
\Return $\{A^1, \dots, A^N\}$
\end{algorithm}

\subsection{Neural Network Design for Allocator}
\label{subsec:neural_network_design}

Our allocator is parameterized by a lightweight feed-forward neural network, which predicts a per-user scale factor that determines how much of their demand should be granted at each step. Formally, the model takes multiple input signals and outputs a nonnegative scalar multiplier applied to the demand vector.

At each step $k$, the network receives as input the step ratio $r_k=\tfrac{k}{N}$, the current demand vector $d_k \in \mathbb{R}^m$, the remaining capacity vector $c_k \in \mathbb{R}^m$, and the previous allocation matrix $A^{k-1} \in \mathbb{R}^{(k-1)\times m}$. These are concatenated into a single feature vector $x_k = [\,r_k ; d_k ; c_k ; A^{k-1}\,] \in \mathbb{R}^{1+(k+1)m}$.

The allocator (formally \textbf{FairUtil} as detailed in Algorithm~\ref{alg:fair_util}) is implemented as a two-layer MLP with ReLU activations between hidden layers. The final layer uses a Softplus activation to produce a nonnegative scalar:
\begin{equation}
    s_k = \text{Softplus}(g(x_k)).
\end{equation}

The predicted scalar $s_k$ is multiplied with the demand vector $d_k$ to form a raw allocation:
\begin{equation}
    \tilde{a}^{k} = s_k \cdot d_k.
\end{equation}
This parameterization explicitly implements \textit{non-wastefulness},
significantly simplifying the target quantity for learning.

The allocation is then capped by the leftover vector:
\begin{equation}
    a^{k} = \min(\tilde{a}^{k}, c_k),
\end{equation}
where the minimum is applied element-wise across resources. This guarantees allocations never exceed currently available resources.

At each step $k$, the resulting allocation $A^{k}$ is recorded and the leftover vector is updated as
\begin{equation}
    c_{k+1} = c_k - a^{k}.
\end{equation}
This sequential process continues until all $N$ users in the window have been processed.

\begin{figure*}[t]
\centering
\includegraphics[width=.32\linewidth]{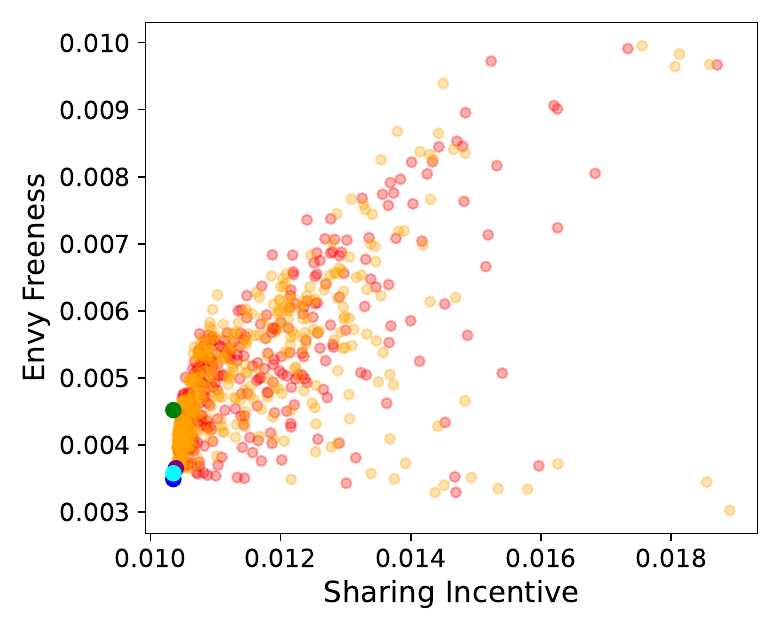}
\includegraphics[width=.32\linewidth]{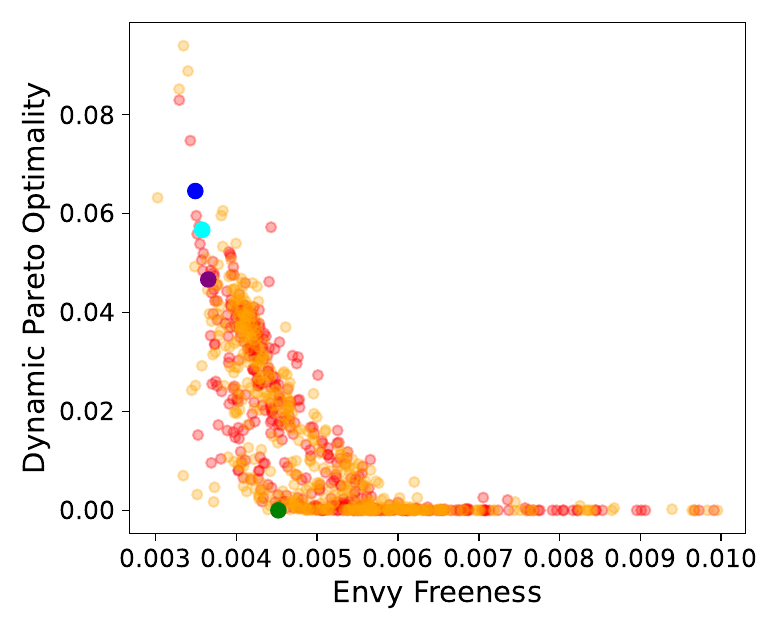}
\includegraphics[width=.32\linewidth]{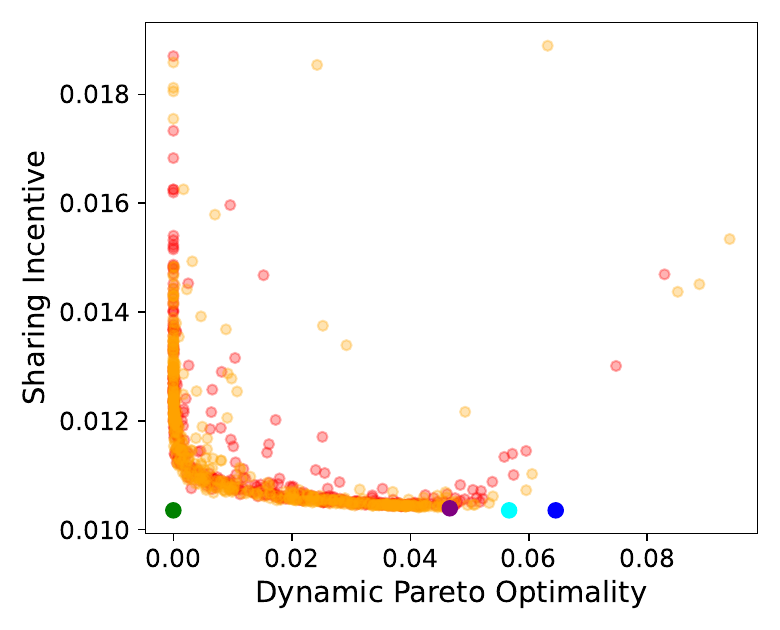}

\vspace{0.75em}

\includegraphics[width=.32\linewidth]{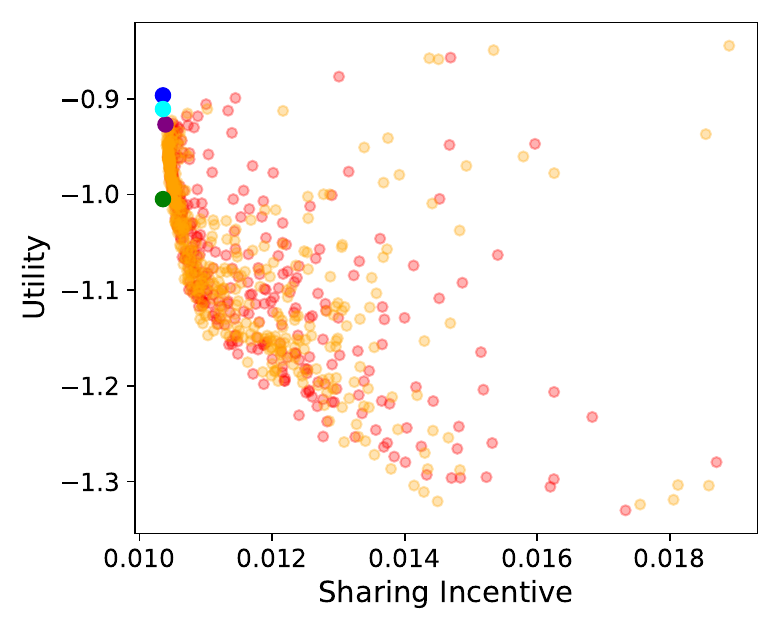}
\includegraphics[width=.32\linewidth]{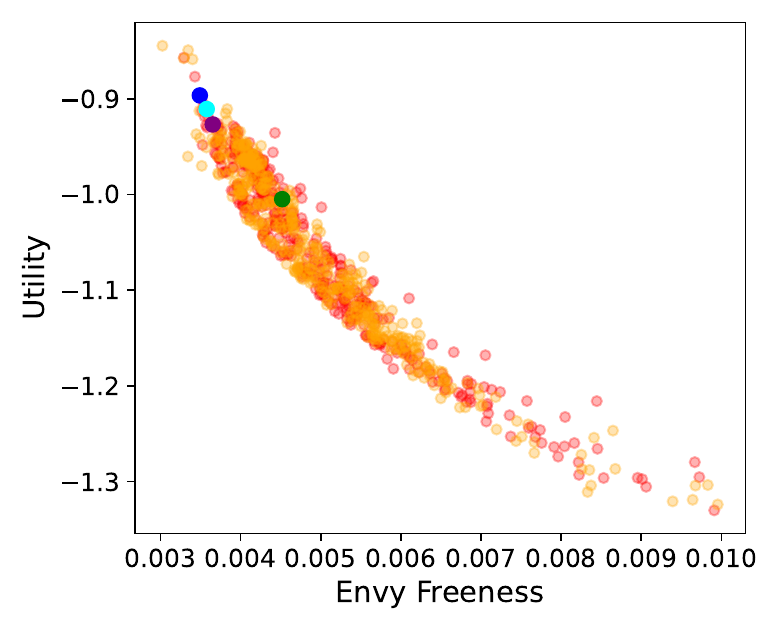}
\includegraphics[width=.32\linewidth]{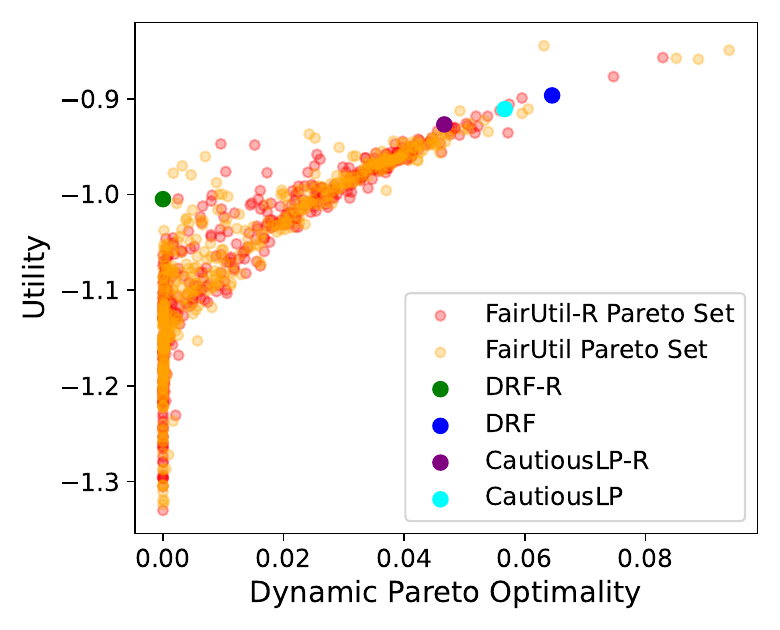}

\caption{Results on Alibaba Cluster Dataset (\texttt{openb\_pod\_list\_cpu100.csv})}
\label{fig:alibaba_cpu}
\end{figure*}

\begin{figure*}[t]
\centering
\includegraphics[width=.32\linewidth]{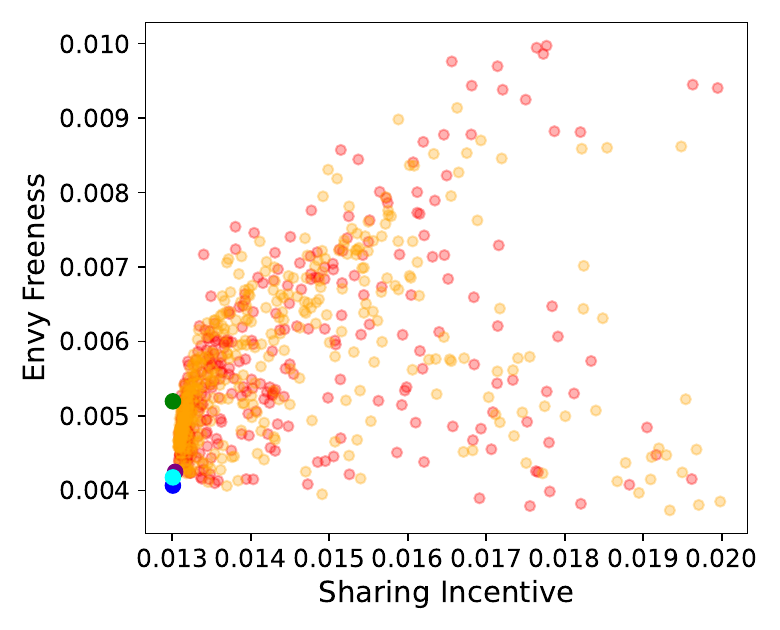}
\includegraphics[width=.32\linewidth]{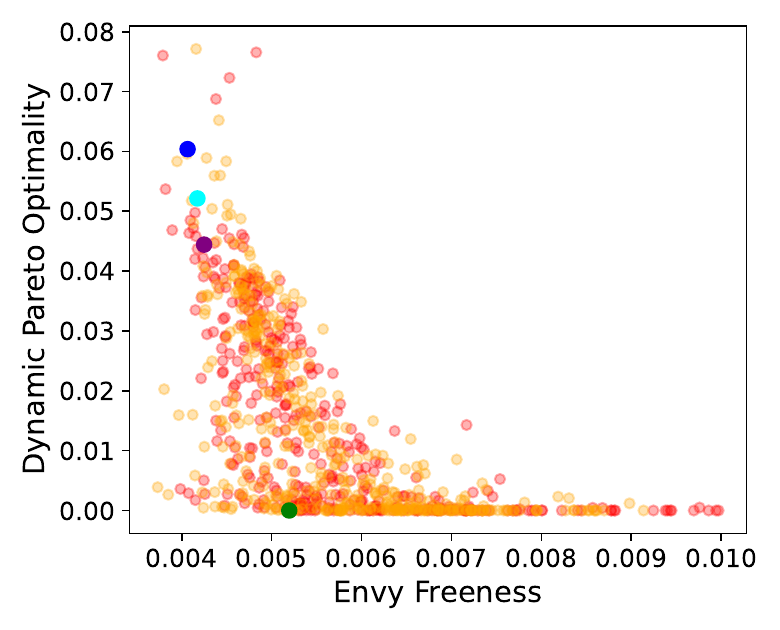}
\includegraphics[width=.32\linewidth]{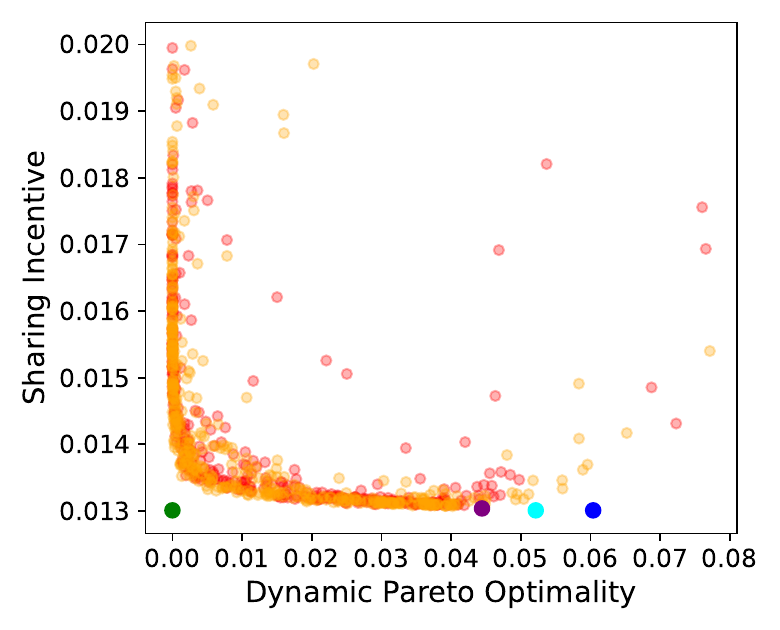}

\vspace{0.75em}

\includegraphics[width=.32\linewidth]{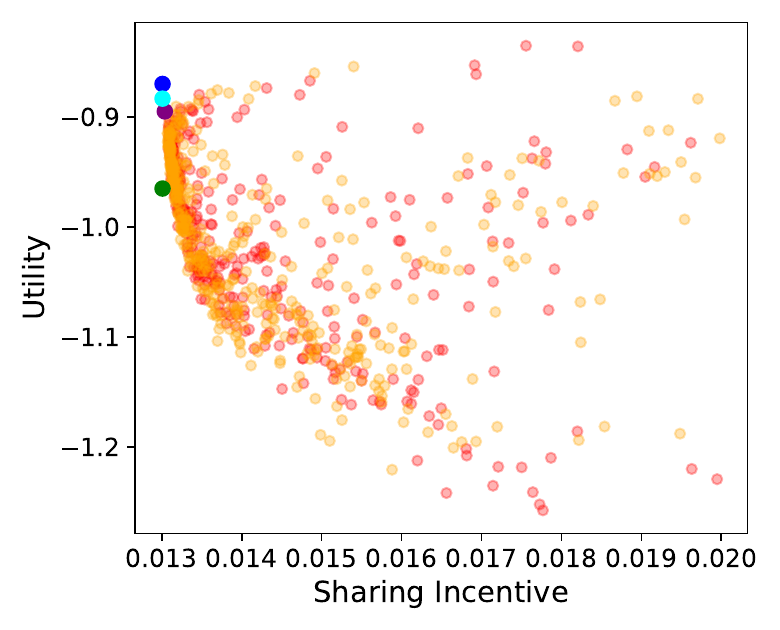}
\includegraphics[width=.32\linewidth]{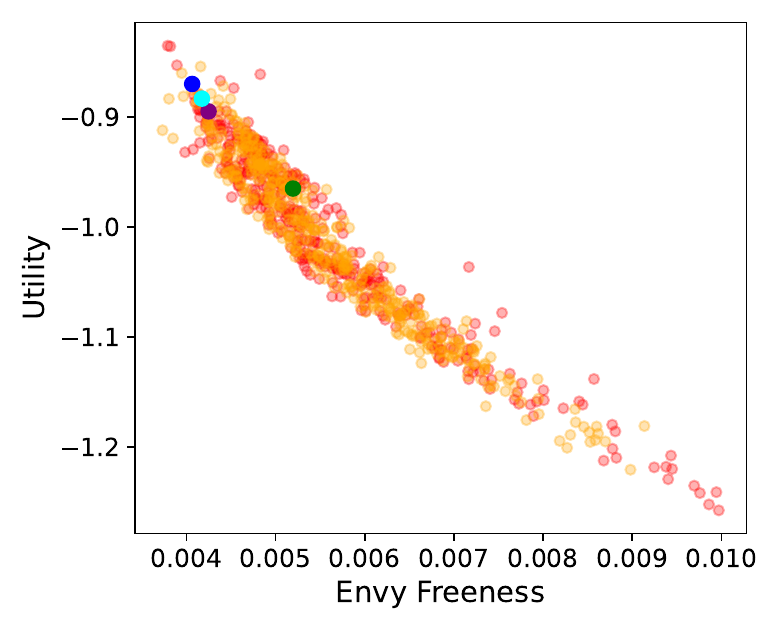}
\includegraphics[width=.32\linewidth]{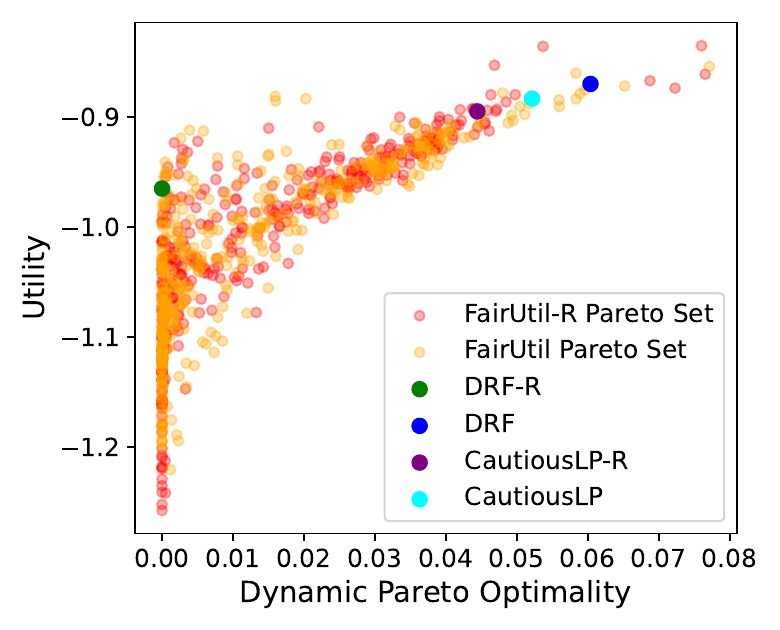}

\caption{Results on Alibaba Cluster Dataset (\texttt{openb\_pod\_list\_gpushare20.csv})}
\label{fig:alibaba_gpu}
\end{figure*}

\begin{figure*}[t]
\centering
\includegraphics[width=.32\linewidth]{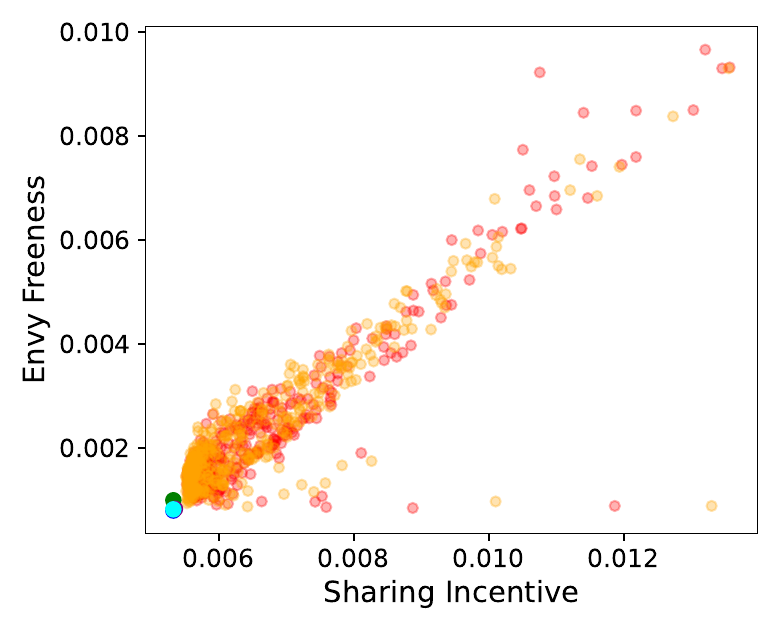}
\includegraphics[width=.32\linewidth]{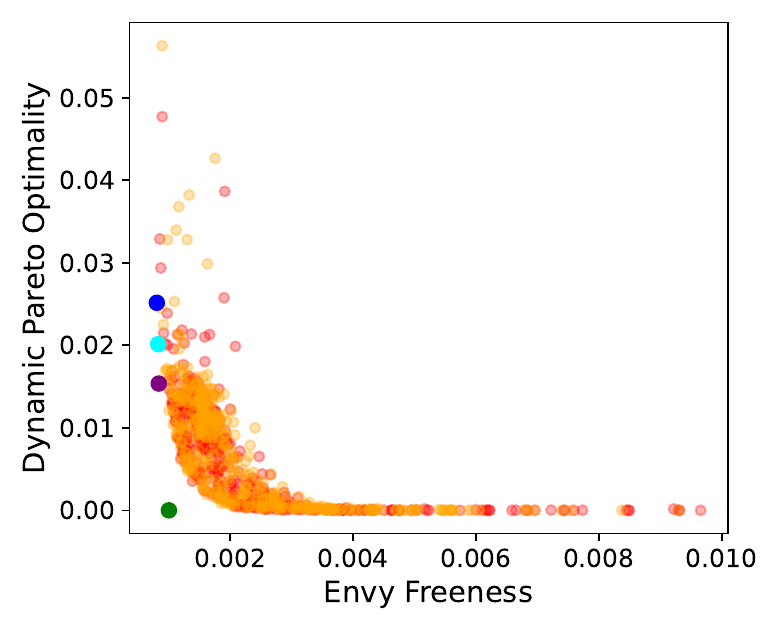}
\includegraphics[width=.32\linewidth]{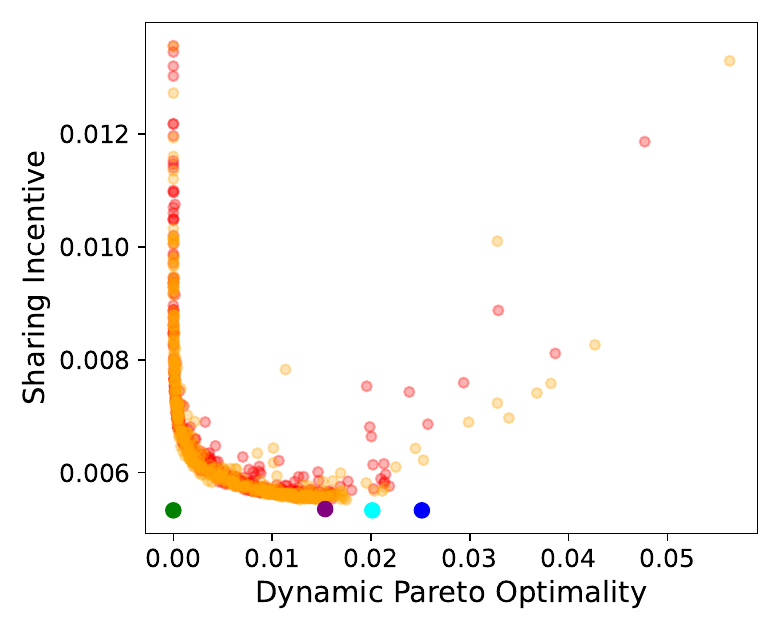}

\vspace{0.75em}

\includegraphics[width=.32\linewidth]{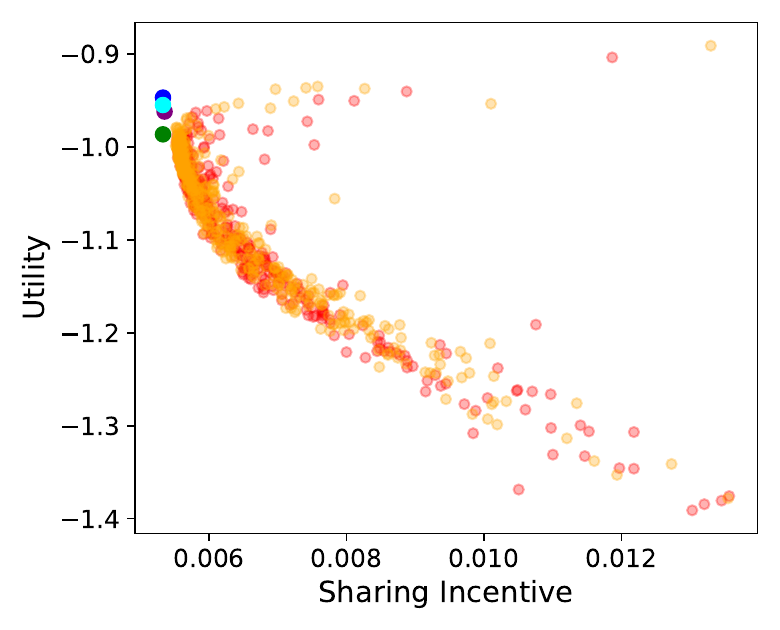}
\includegraphics[width=.32\linewidth]{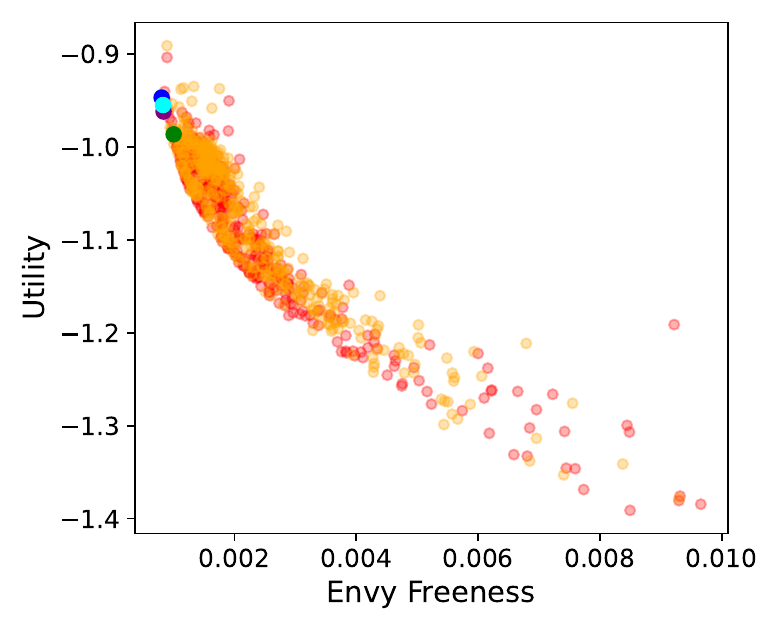}
\includegraphics[width=.32\linewidth]{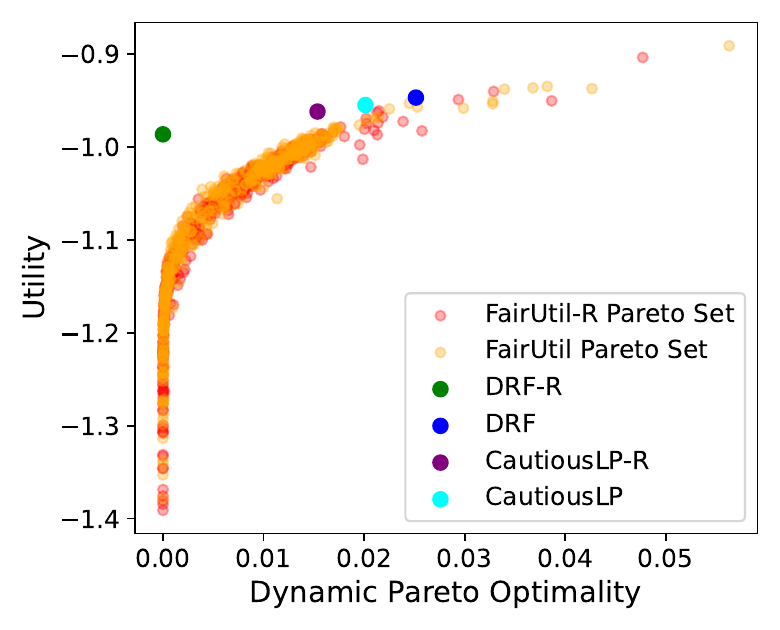}

\caption{Results on Azure dataset}
\label{fig:azure}
\end{figure*}

\begin{algorithm}[t]
\caption{FairUtil-R allocator}
\label{alg:fair_util_r}
\KwIn{demands $\{d_k\}_{k \in [N]}$, total capacity $c_1 \in \mathbb{R}^m$}
\KwOut{allocations $\{A^1,\dots,A^N\}$}

\For{$k = 1$ \KwTo $N$}{
    $r_k \gets \tfrac{k}{N}$\;
    $s_{1:k} \gets \text{Softplus}\big(g([\,r_k;\, d_{1:k};\, A^{k-1};\, c_k\,])\big)$\;
    $\Delta\tilde{A}^k \gets [s_1 d_1, \ldots, s_k d_k]^\top$ \;
    $q \gets \sum_{i=1}^k \Delta\tilde{A}^k_i$ (proposed increment)\;
    $\alpha \gets \min\!\big(\mathbf{1},\, c_k / (q + \varepsilon)\big)$ \ (elementwise)\; 
    $\Delta A^k \gets [\alpha_1 \Delta\tilde{A}^k_1, \ldots, \alpha_k \Delta\tilde{A}^k_k]^\top$\;
    $A^k \gets [A^{k-1}, \zero] + \Delta A^k$\;
    $c_{k+1} \gets c_k - \sum_{i=1}^k \Delta A^k_i$\;
}
\Return $\{A^1,\dots,A^N\}$
\end{algorithm}

\subsection{Replenishing Allocations for Previous Users}

FairUtil assigns resources exclusively to the newly arrived user at step $k$, while freezing allocations made to earlier users. To further enhance flexibility and fairness, we introduce a variant, \textbf{FairUtil-R} which allows allocations to previous users $1, \dots, k-1$ to be incrementally replenished when additional resources remain available.
The method is detailed in Algorithm~\ref{alg:fair_util_r}.

At each step $k$, FairUtil-R receives as input the demands $d_{1:k}$, the previous allocations $A^{k-1}$ for the active users, the step ratio $r_k$, and the leftover capacity $c_k$. For each active user $i \leq k$, the model proposes a nonnegative \emph{increment} vector
\begin{equation}
\Delta A_i^k = \delta_i^k \cdot d_i, \quad \delta_i^k \geq 0,
\end{equation}
where $\delta_i^k$ is predicted by a small MLP applied row-wise. These increments are parallel to the demand vectors, ensuring proportionality to user needs. To preserve feasibility, the deltas are globally capped so that their total does not exceed the available leftover $c_k$:
\begin{equation}
\Delta A^k = \{\Delta A_1^k, \dots, \Delta A_{k}^k\}, \quad
\sum_{i=1}^{k} \Delta A_i^k \leq c_k.
\end{equation}

The allocations for the first $k-1$ users are updated monotonically:
\begin{equation}
A_i^k = A_i^{k-1} + \Delta A_i^k, \quad \forall i < k,
\end{equation}
while the new user $k$ is allocated $\Delta A_{k}^k$. The leftover capacity is then updated as
\begin{equation}
c_{k+1} = c_k - \sum_{i=1}^{k} \Delta A_i^k.
\end{equation}

\section{Experiments}
\subsection{Datasets} 
We evaluate our method on two real-world cluster workload traces: 
\begin{itemize}
\vspace{-0.5em}
\setlength\itemsep{0.5em}
\setlength\parskip{0pt}
    \item \textbf{Alibaba Cluster Dataset (v2023).} The Alibaba Cluster Dataset is a large public repository containing datacenter traces from multiple years \citep{alibaba2023clusterdata}. We specifically use the 2023 release, which records pod-level demands across heterogeneous resources. In our experiments, we extract three resource dimensions: \texttt{CPU}, \texttt{memory}, and \texttt{GPU}. We focus on two representative CSV files: \texttt{openb\_pod\_list\_cpu100.csv} (7{,}853 entries) and \texttt{openb\_pod\_list\_gpushare20.csv} (8{,}152 entries).
    \item \textbf{Microsoft Azure Traces (Packing 2020).}
    The Microsoft Azure traces comprise a relatively large collection of scheduling and allocation data from production clusters. 
    For tractability, we focus on the \texttt{AzureTracesForPacking2020} dataset, which is provided as a SQLite file. 
    We preprocess the \texttt{vmType} table and extract three resource dimensions \texttt{memory}, \texttt{ssd}, and \texttt{nic} to form a multi-resource allocation setting. 
    The processed dataset contains $4{,}619$ entries in total.
\end{itemize}

During preprocessing, we first sort all entries (i.e., demands) by their start time. We then construct sliding windows of size 10 over consecutive demands, resulting in $(\text{number of entries} - \text{window size})$ windows. Finally, these windows are randomly grouped into batches for training.

\subsection{Experiment Setting}
Unless otherwise specified, we set $N=10$, meaning that allocations are computed over windows of 10 sequential users. This choice allows sufficient dynamic interactions to be captured.
We defer results for other values of $N$ to appendix.

We experimented on the following methods:
\vspace{-0.5em}
\begin{itemize}
\setlength\itemsep{0.5em}
\setlength\parskip{0pt}
    \item \textbf{FairUtil}: our supervised fairness–utility framework.
    \item \textbf{FairUtil-R}: replenishment-enabled extension of FairUtil.
    \item \textbf{DRF}: Dynamic Dominant Resource Fairness~\citep{ghodsi2011dominant}.
    \item \textbf{DRF-R}: replenishment-enabled extension of DRF~\citep{kash2013no}.
    \item \textbf{CautiousLP}: A linear programming-based mechanism introduced by~\citet{kash2013no}, designed to strictly enforce SI, EF, and CDPO (a relaxed variant of DPO).
    \item \textbf{CautiousLP-R}: replenishment-enabled extension of CautiousLP.
\end{itemize}

We present six plots for each dataset. 
The top row illustrates the pairwise trade-offs between fairness metrics/losses (SI vs.\ EF, EF vs.\ DPO, and DPO vs.\ SI), 
which allows us to examine the inherent tensions among fairness notions. 
These visualizations also serve to verify that our method does not sacrifice fairness, 
and in many cases is able to maintain comparable fairness with existing baselines. 

The bottom row presents fairness metrics against utility (SI vs.\ UT, EF vs.\ UT, and DPO vs.\ UT), 
highlighting the central fairness--utility trade-off that motivates our work.
We flipped the utility so that the ideal direction points towards the origin.

\subsection{Implementation Details} 
All experiments are conducted on a single NVIDIA RTX 2080Ti GPU with 11GB of memory. Since our evaluation is based on the Pareto set rather than a single point, validation plays only a minor role. Nonetheless, we tune hyperparameters once and then keep them fixed across all experiments for consistency. Stable performance is observed for batch sizes between 16 and 256, and we set the batch size to 128. A learning rate sweep over $[1\times 10^{-3}, 1.6\times 10^{-2}]$ shows reliable convergence, leading us to fix the learning rate at $8\times 10^{-3}$. The number of training epochs is set to 3 in all experiments. With these hyperparameters fixed, we train our models and report the resulting Pareto set by varying only the fairness loss weights.

\paragraph{Grid Enumeration for Loss Weights} 
Training our model requires choosing the relative weights $(\lambda_{\text{SI}}, \lambda_{\text{EF}}, \lambda_{\text{DPO}})$ for the stepwise fairness losses. This selection is non-trivial. To systematically explore this trade-off, we adopt a logarithmic grid search.

We fix $\lambda_{\text{DPO}} = 1$ and sweep the remaining weights. 
Specifically, $\lambda_{\text{SI}}$ is varied over 10 log-spaced values in the range $[0.5, 2\times 10^4]$, 
while $\lambda_{\text{EF}}$ is varied across 7 log-spaced values in $[0.1, 1000]$. 
This produces a pool of $10 \times 7 = 70$ candidate weight pairs. 
Each triplet $(\lambda_{\text{SI}}, \lambda_{\text{EF}}$,$\lambda_{\text{DPO}})$
defines one model instance. Evaluating all instances yields the Pareto set of fairness--utility trade-offs.

\subsection{Experimental Results}
\vspace{-0.5em}
Our experimental results on the Alibaba Cluster Dataset (Figure~\ref{fig:alibaba_cpu}, Figure~\ref{fig:alibaba_gpu}) and Microsoft Azure dataset (Figure~\ref{fig:azure}) reveal several consistent patterns. First, SI and EF tend to be positively correlated: allocations that achieve low SI loss almost always achieve low EF loss as well. This indicates that SI and EF do not constitute a meaningful trade-off, and drawing a Pareto frontier in this case would be uninformative. For this reason, we intentionally plot all runs without connecting a frontier, or equivalently Pareto set. In contrast, DPO shows a pronounced tension with both SI and EF: improvements in DPO are consistently accompanied by increases in SI or EF, indicating the inherent conflict between efficiency and fairness objectives. This phenomenon aligns with the impossibility result of \citet{kash2013no}, which proved that EF and DPO cannot be simultaneously guaranteed in dynamic settings.

Beyond these fairness–fairness trade-offs, we also observe clear interactions with utility. SI and EF are generally negatively correlated with utility, reflecting the cost of enforcing stricter fairness. By contrast, DPO is correlated with utility, since reducing DPO reduces wasted resources.

Theorem \ref{theorem:re_dpo_utility} shows that the cumulative average utility is lower-bounded by a simple function of the stepwise DPO losses. In other words, if we enforce DPO by minimizing the weighted loss $\sum_{k=1}^N \ell_{\mathrm{DPO}}^k/k$, then the allocation policy must achieve high cumulative utility in the sense captured by $\sum_{k=1}^N U_k$.
\begin{theorem}
\label{theorem:re_dpo_utility}
\begin{align}
    let \ U_k \;&=\; \frac{1}{k}\sum_{i=1}^k x_i, \\
    \sum_{k=1}^N U_k \;
    &\ge\; 1 - \sum_{k=1}^N \frac{\ell_{\mathrm{DPO}}^k}{k}.
\end{align}
\end{theorem}
This makes the DPO loss a natural surrogate for utility, implicitly guiding the model toward higher throughput even though utility is not explicitly optimized in training. These findings confirm our initial intuition from the DPO's definition that utility can be implicitly captured through some fairness metrics. We defer the proof of theorem \ref{theorem:re_dpo_utility} to appendix \ref{sec:re_dpo_utility}

\textbf{For completeness, the supplementary material provides extra experimental results that directly add utility to the training objective in \eqref{eq:fair-loss}.}

Most importantly, our experiments demonstrate a striking fairness–utility trade-off. Enforcing stricter fairness constraints reduces overall throughput, a limitation clearly observed in classical mechanisms such as DRF and Cautious LP: while these approaches provide strong fairness guarantees, they consistently under-utilize available resources and suffer from degraded utility. By contrast, our neural allocation mechanism achieves fairness levels comparable to these baselines while delivering substantially higher utility. This marks a critical advantage: rather than \textit{strictly} enforcing fairness at the expense of efficiency, our method learns to balance the two in a principled way. 

Interestingly, we also observe that \textbf{FairUtil} performs comparably to \textbf{FairUtil-R} in most cases, despite the lack of replenishment. This suggests that the proposed learning mechanism adapts effectively to sequential arrivals, and that replenishment may not always be critical to achieving strong fairness–utility trade-offs in practice. 

Overall, our results not only confirm the theoretical fairness tensions in dynamic fair division but also provides a practical solution that achieves significant utility improvements without marked compromise of fairness.

\section{Conclusion}
We presented \textbf{FairUtil}, a neural allocation mechanism for dynamic multi-resource allocation that directly optimizes fairness–utility tradeoffs through differentiable fairness losses and sequential rollout. Our method consistently achieves substantially higher utility while maintaining fairness at levels comparable to established baselines. Experiments on real-world traces validate both the theoretical tensions in dynamic fair division and the practical strength of our approach, revealing tradeoffs that extend beyond prior theoretical work and offer an interpretable and effective solution in practice. For future work, it would be useful to move beyond non-wasteful allocations and study richer utility models. We can also incorporate additional constraints, such as priority policies and timeline requirements. Finally, we can improve robustness to workload shifts by adapting policies across traces and demand distributions.

\newpage
\bibliography{uai2026-template}

\newpage

\onecolumn

\title{Supplementary Material}
\maketitle
\appendix
\section{Relationship between DPO and Utility}
\label{sec:re_dpo_utility}
Assume each user $i\le k$ has a demand vector $d_i\in\mathbb{R}_{\ge 0}^m$ that is
\emph{dominant-normalized}:
\begin{equation}
\max_{r\in[m]} d_{ir} \;=\; 1 .
\label{eq:dominant-norm}
\end{equation}
Under non-wastefulness (proportional allocations), we write $A_i = x_i\, d_i$ with $x_i\ge 0$, so the per-user utility reduces to the scale factor (ignoring the case where $d_{ir}=0$):
\begin{equation}
u_i \;=\; \min_{r:\,d_{ir}>0}\frac{A_{ir}}{d_{ir}} \;=\; x_i,
\qquad
U_k \;=\; \frac{1}{k}\sum_{i=1}^k x_i .
\label{eq:Uk-def}
\end{equation}
Let $S_{k,r}=\sum_{i=1}^k A_{ir}=\sum_{i=1}^k x_i d_{ir}$ be total allocation of resource $r$, and $S_k^{\max}=\max_r S_{k,r}$ the  dominant resource usage. The DPO loss we optimize is
\begin{equation}
\ell_{\mathrm{DPO}}^k \;=\; \big(\tfrac{k}{N}-S_k^{\max}\big)_+ .
\label{eq:DPO-loss}
\end{equation}

From \eqref{eq:dominant-norm} we have $0\le d_{ir}\le 1$, hence for every $r$,
\begin{equation}
S_{k,r} \;=\; \sum_{i=1}^k x_i d_{ir} \;\le\; \sum_{i=1}^k x_i
\quad\Longrightarrow\quad
S_k^{\max}\;\le\;\sum_{i=1}^k x_i .
\label{eq:Smax-upper}
\end{equation}
Combining \eqref{eq:Uk-def} and \eqref{eq:Smax-upper} yields the stepwise lower bound
\begin{equation}
U_k \;\ge\; \frac{1}{k}\,S_k^{\max}.
\label{eq:Uk-lb-by-Smax}
\end{equation}
From \eqref{eq:DPO-loss} we also have $S_k^{\max}\ge \tfrac{k}{N}-\ell_{\mathrm{DPO}}^k$, thus
\begin{equation}
U_k \;\ge\; \frac{1}{N}-\frac{1}{k}\,\ell_{\mathrm{DPO}}^k .
\label{eq:Uk-vs-DPO}
\end{equation}
Summing \eqref{eq:Uk-vs-DPO} over $k=1,\dots,N$ gives
\begin{equation}
\sum_{k=1}^N U_k \;\ge\; 1 - \sum_{k=1}^N \frac{\ell_{\mathrm{DPO}}^k}{k}.
\label{eq:sum-Uk}
\end{equation}

Under proportional (non-wasteful) allocations with dominant-normalized demands \eqref{eq:dominant-norm}, decreasing the DPO loss \eqref{eq:DPO-loss} \emph{monotonically tightens a lower bound} on aggregate utility via \eqref{eq:Uk-vs-DPO}–\eqref{eq:sum-Uk}. Equivalently, the DPO loss is \emph{negatively related with utility}: reducing $\ell_{\mathrm{DPO}}^k$ raises the utility even though utility is not explicitly optimized during training.

\section{Pearson Coefficient}
To better understand the correlation among fairness metrics and their relationship with utility, we compute the \textit{Pearson coefficients} between all pairwise metrics 
(\textsc{SI}, \textsc{EF}, \textsc{DPO}, \textsc{UT}) across all experimental runs as shown in Table~\ref{tab:pearson_corr}.
\begin{table}[t]
\centering
\caption{Pairwise Pearson correlation coefficients among fairness metrics (SI, EF, DPO) and utility (UT) across datasets. 
Positive correlations indicate alignment, while negative values reflect trade-offs.}
\label{tab:pearson_corr}
\resizebox{\linewidth}{!}{
\begin{tabular}{lcccccc}
\toprule
\textbf{Method / Dataset} & (SI, EF) & (EF, DPO) & (DPO, SI) & (SI, UT) & (EF, UT) & (DPO, UT) \\
\midrule
\textbf{FairUtil-R (CPU100)} & 0.375 & -0.676 & -0.570 & -0.545 & -0.890 & 0.914 \\
\textbf{FairUtil (CPU100)}   & 0.232 & -0.800 & -0.520 & -0.383 & -0.941 & 0.945 \\
\midrule
\textbf{FairUtil-R (GPUShare20)} & 0.478 & -0.618 & -0.476 & -0.469 & -0.955 & 0.788 \\
\textbf{FairUtil (GPUShare20)}   & 0.308 & -0.636 & -0.523 & -0.339 & -0.965 & 0.780 \\
\midrule
\textbf{FairUtil-R (Azure)} & 0.927 & -0.561 & -0.424 & -0.858 & -0.911 & 0.802 \\
\textbf{FairUtil (Azure)}   & 0.886 & -0.601 & -0.369 & -0.816 & -0.925 & 0.813 \\
\bottomrule
\end{tabular}}
\end{table}

\section{Additional Experimental Results}
\subsection{Utility-Aware Training}
\begin{figure*}[t]
\centering
\includegraphics[width=.32\linewidth]{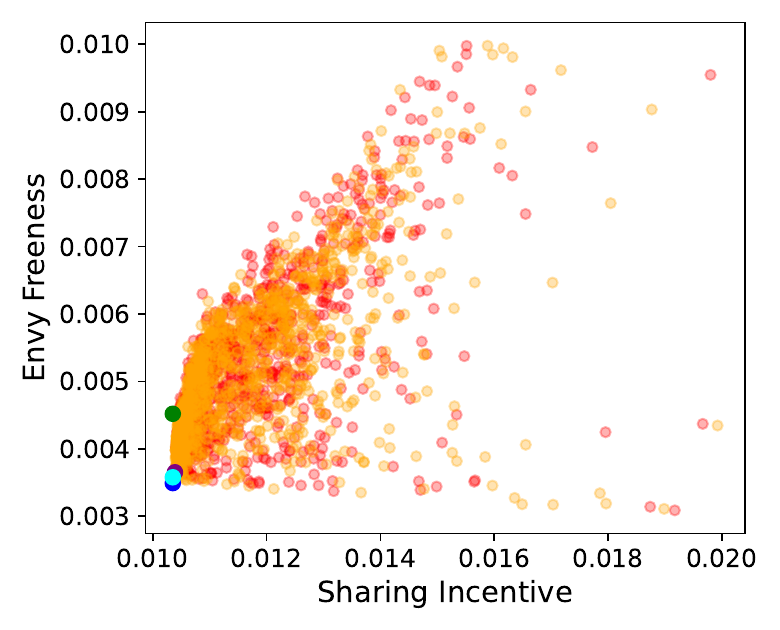}
\includegraphics[width=.32\linewidth]{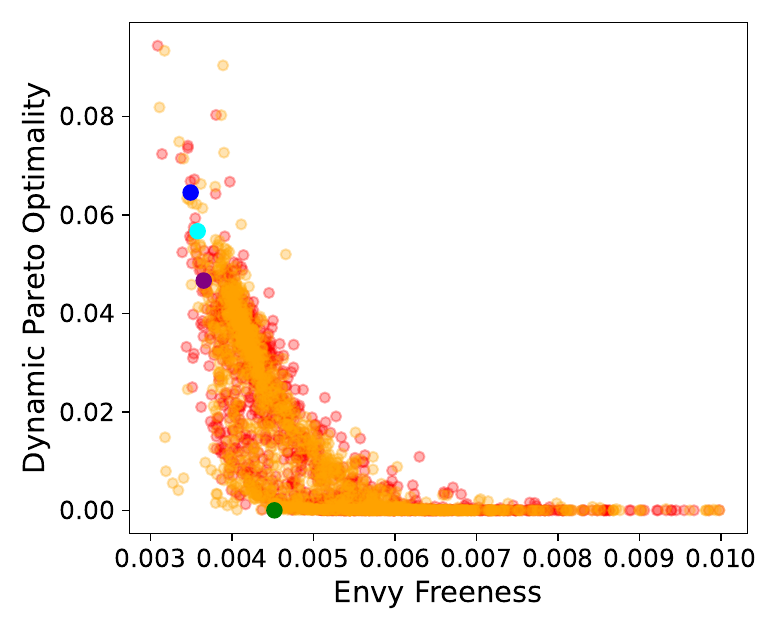}
\includegraphics[width=.32\linewidth]{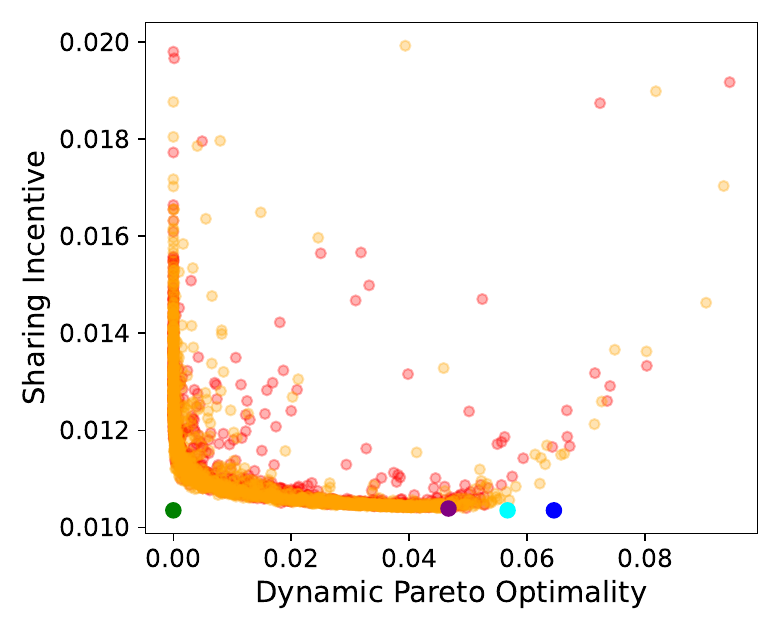}

\vspace{0.75em}

\includegraphics[width=.32\linewidth]{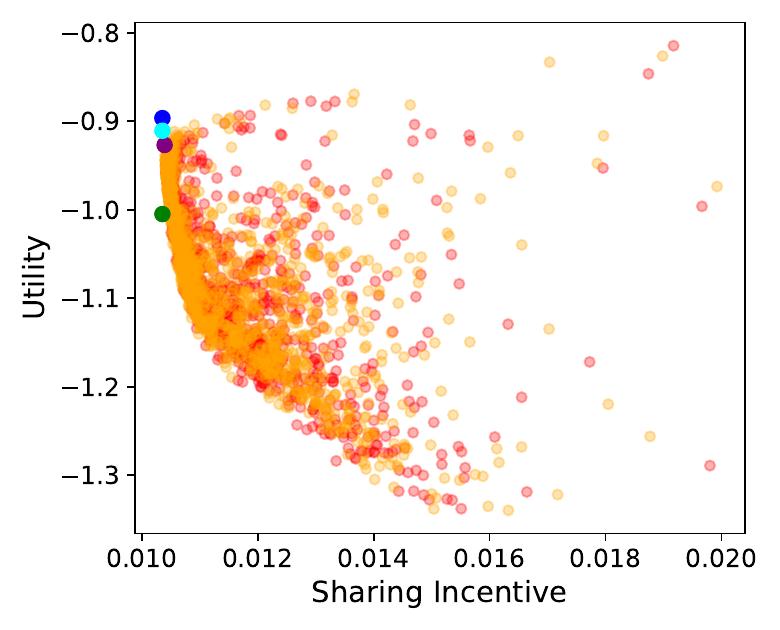}
\includegraphics[width=.32\linewidth]{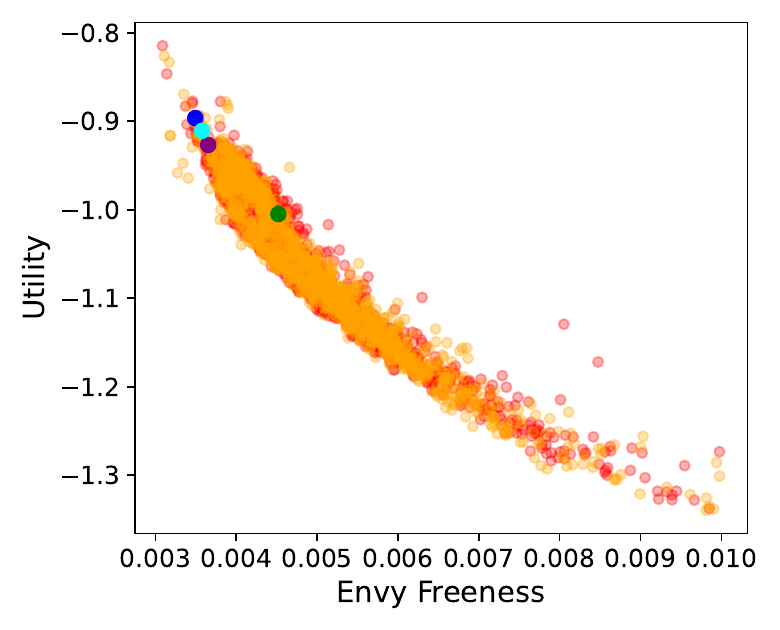}
\includegraphics[width=.32\linewidth]{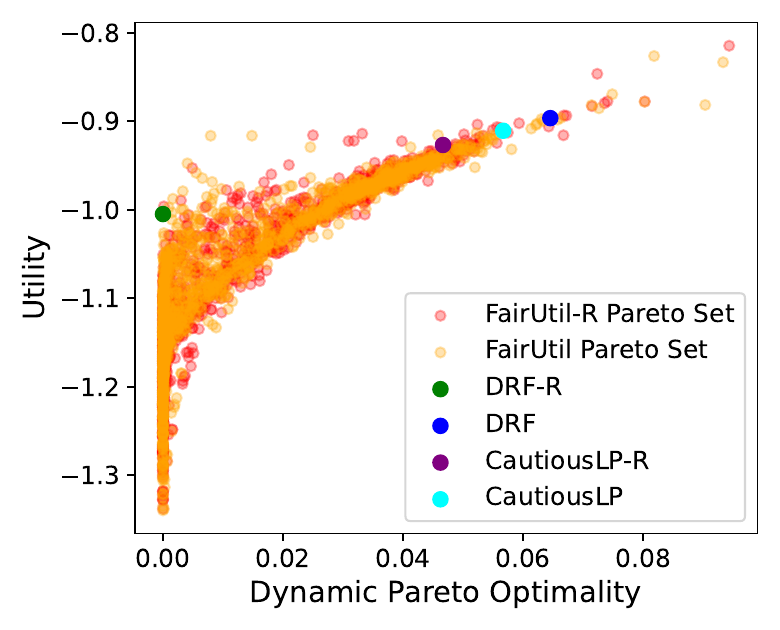}

\caption{Results on Alibaba Cluster Dataset (\texttt{openb\_pod\_list\_cpu100.csv}) with Utility-Aware Training}
\label{fig:appendix_exp_ut_alibaba_cpu}
\end{figure*}

\begin{figure*}[t]
\centering
\includegraphics[width=.32\linewidth]{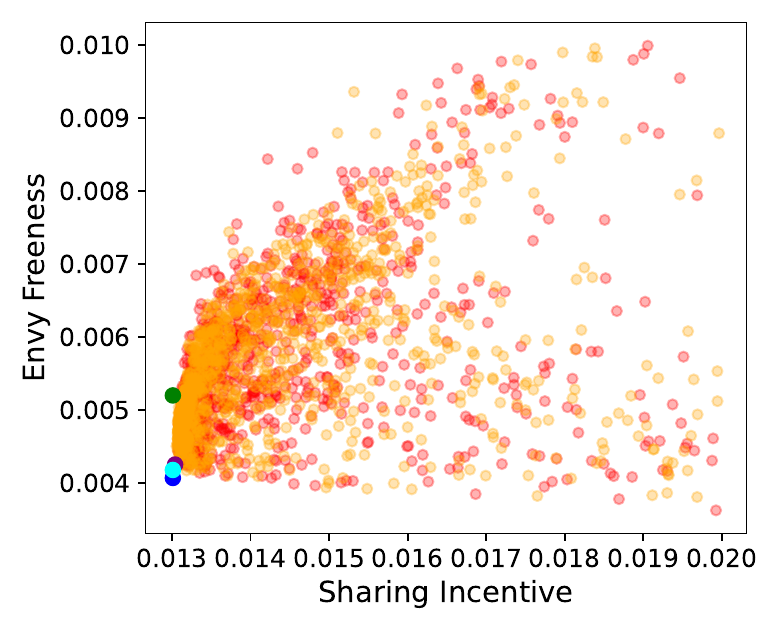}
\includegraphics[width=.32\linewidth]{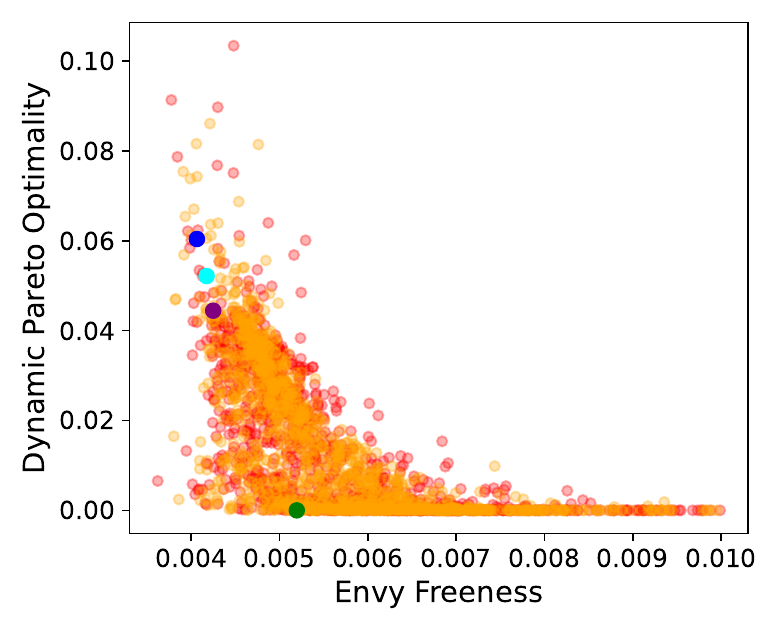}
\includegraphics[width=.32\linewidth]{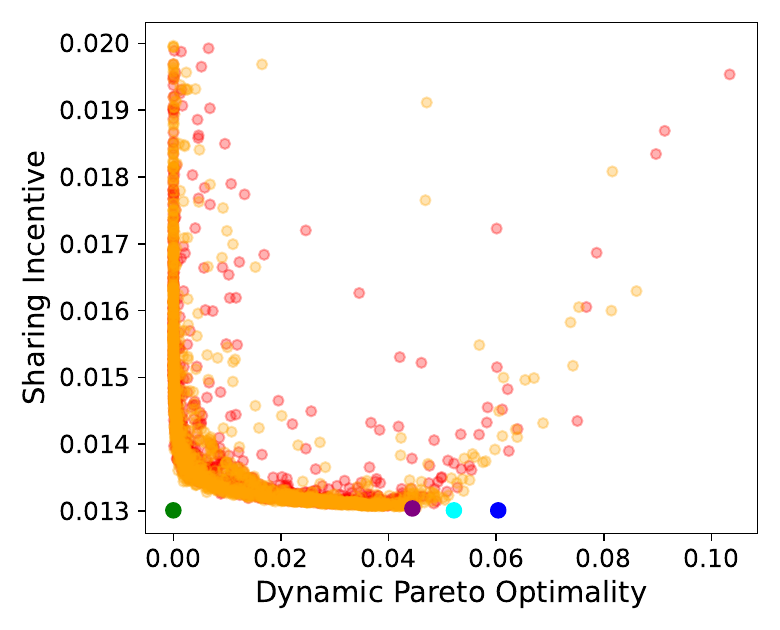}

\vspace{0.75em}

\includegraphics[width=.32\linewidth]{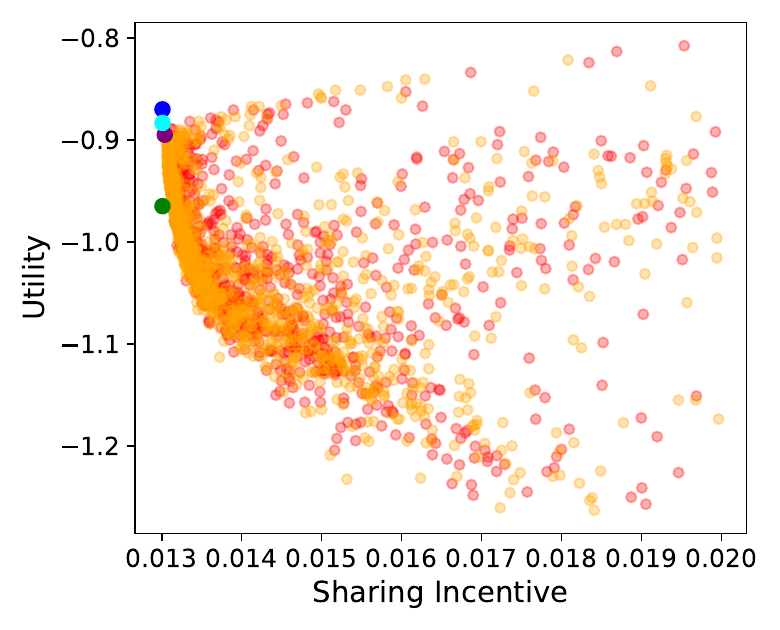}
\includegraphics[width=.32\linewidth]{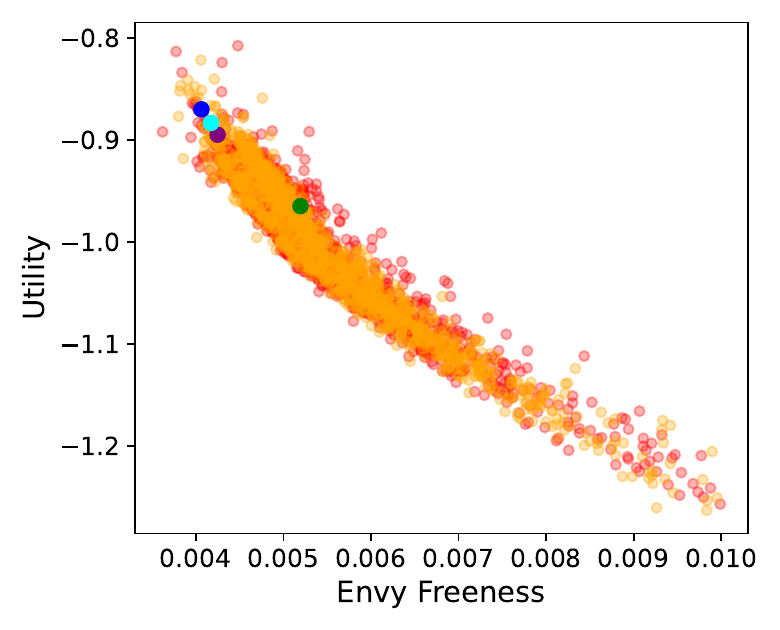}
\includegraphics[width=.32\linewidth]{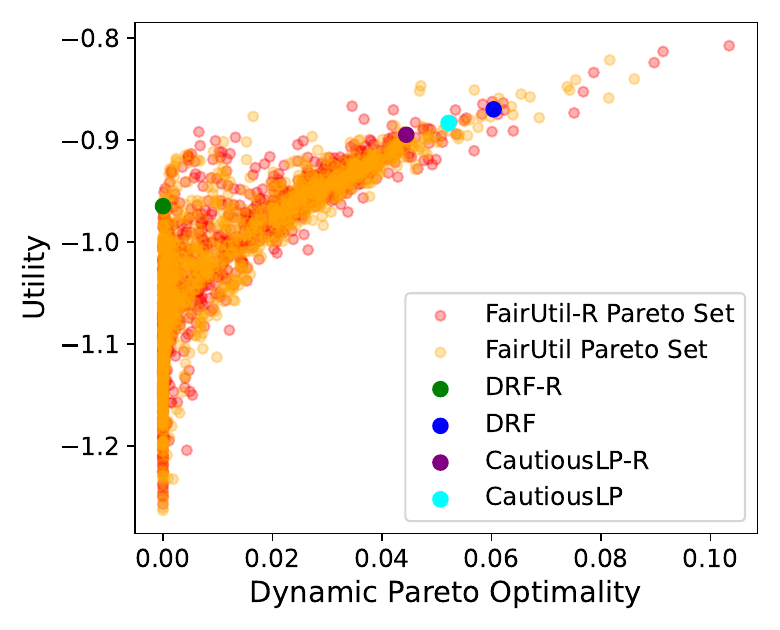}

\caption{Results on Alibaba Cluster Dataset (\texttt{openb\_pod\_list\_gpushare20.csv}) with Utility-Aware Training}
\label{fig:appendix_exp_ut_alibaba_gpu}
\end{figure*}

\begin{figure*}[t]
\centering
\includegraphics[width=.32\linewidth]{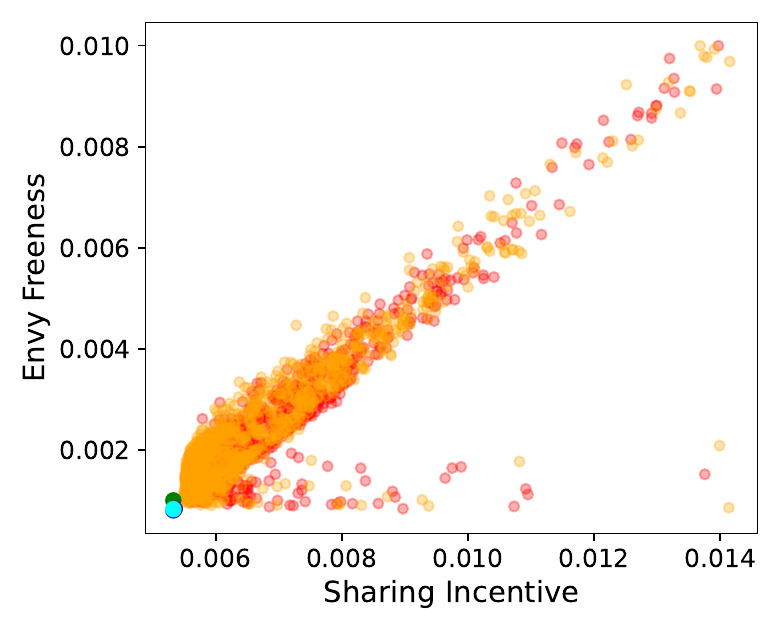}
\includegraphics[width=.32\linewidth]{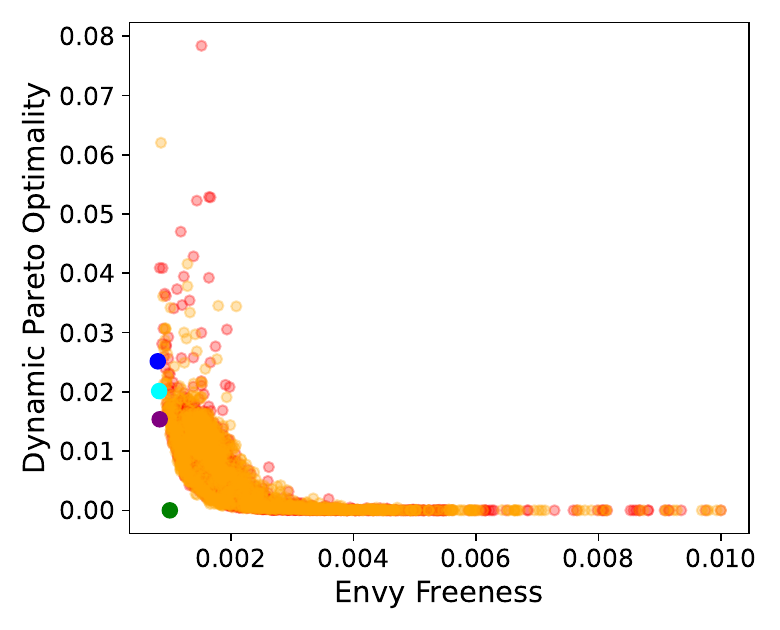}
\includegraphics[width=.32\linewidth]{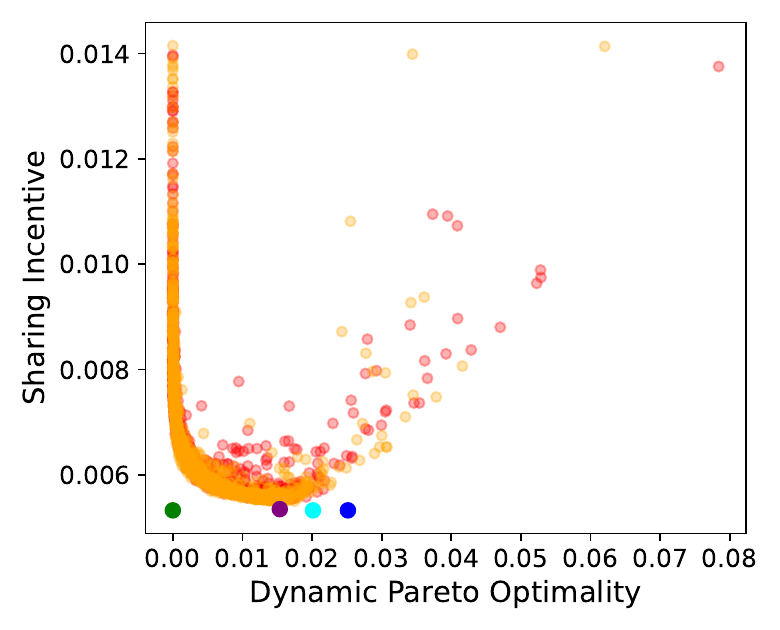}

\vspace{0.75em}

\includegraphics[width=.32\linewidth]{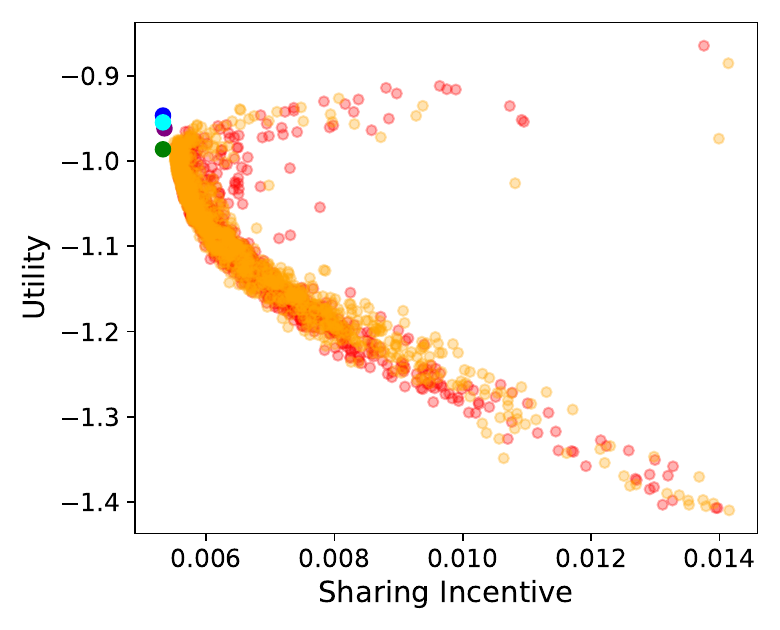}
\includegraphics[width=.32\linewidth]{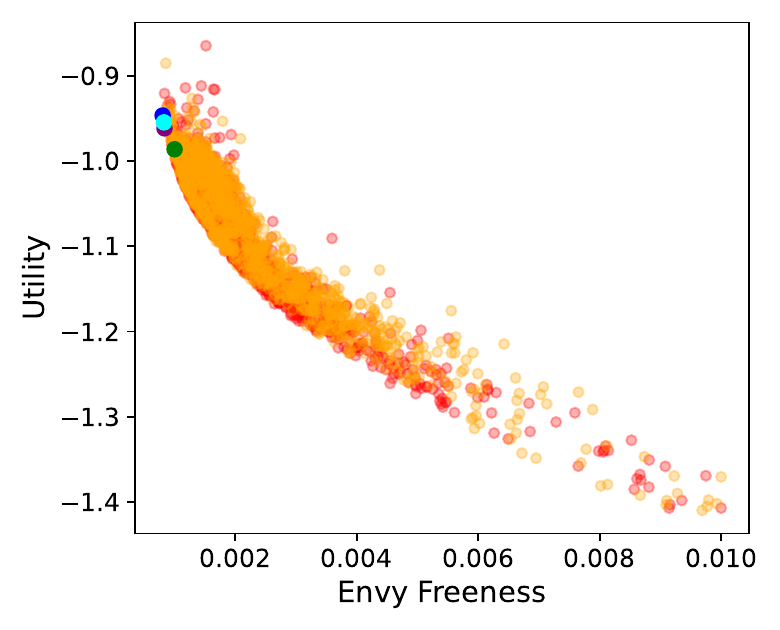}
\includegraphics[width=.32\linewidth]{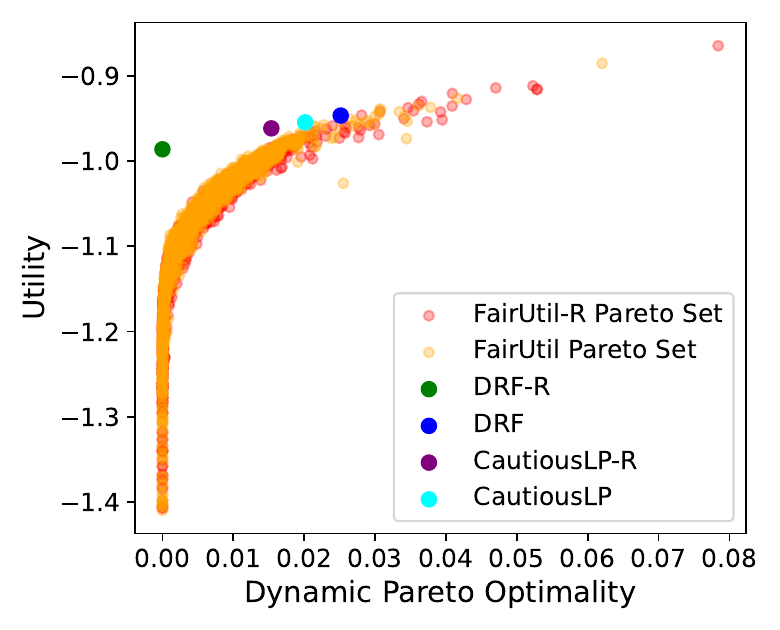}

\caption{Results on Azure dataset with Utility-Aware Training}
\label{fig:appendix_exp_ut_azure}
\end{figure*}

To assess whether explicitly modeling utility improves performance, we extend our training objective with a utility loss term, defined as its negation. 
The corresponding weight $\lambda_{\text{UT}}$ is varied across 3 log-spaced values in $[10^{-4}, 10^{-2}]$. 
The Pareto scatter plots are shown in Figures~\ref{fig:appendix_exp_ut_alibaba_cpu}, \ref{fig:appendix_exp_ut_alibaba_gpu} and \ref{fig:appendix_exp_ut_azure}.
Compared with Figure~\ref{fig:alibaba_cpu}, \ref{fig:alibaba_gpu} and \ref{fig:azure} which do not explicitly include utility in the objective,
this addition turns out to yield minimal improvement,
which aligns with our initial hypothesis that DPO is positively correlated with the negation of utility, 
thus effectively serving as a proxy for utility optimization. 
In other words, minimizing DPO already encourages higher utility, 
confirming that an explicit utility term may not be worth the computational cost in practice (mainly due to model selection) .

\subsection{Larger Window Size}
\begin{figure*}[t]
\centering
\includegraphics[width=.32\linewidth]{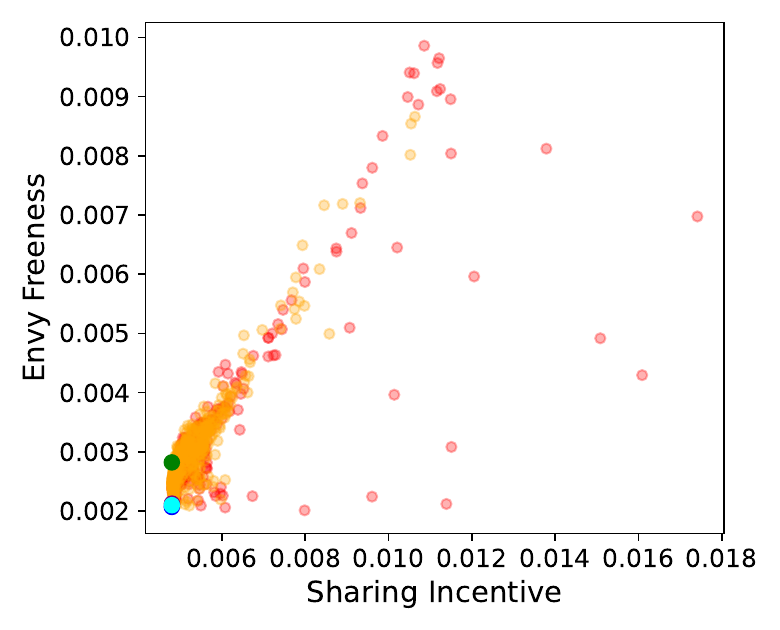}
\includegraphics[width=.32\linewidth]{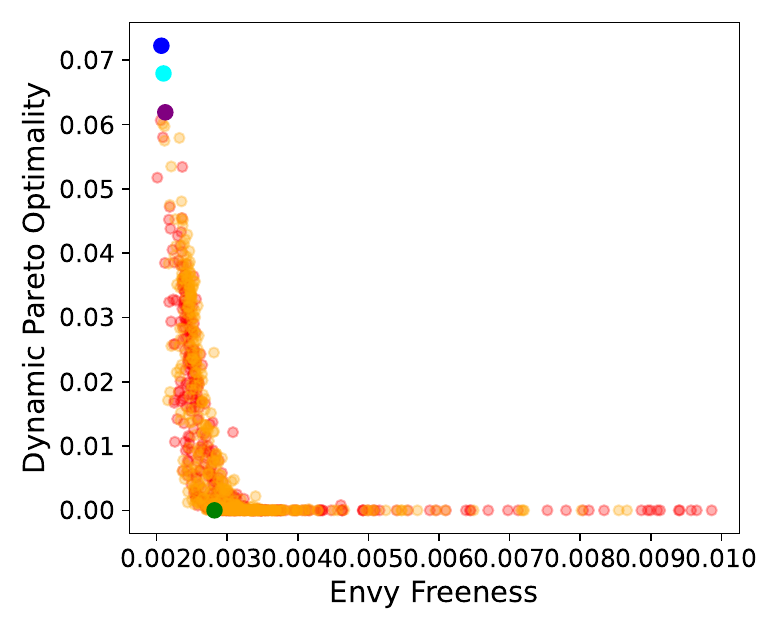}
\includegraphics[width=.32\linewidth]{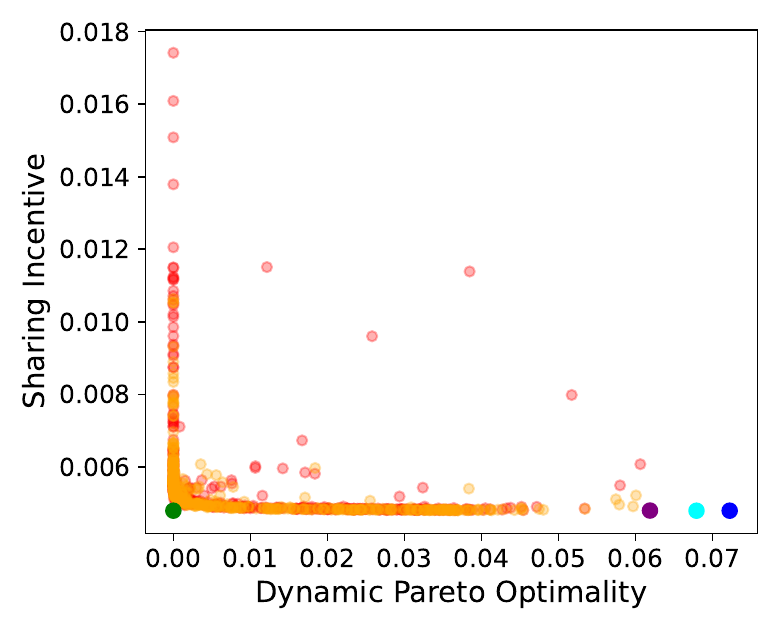}

\vspace{0.75em}

\includegraphics[width=.32\linewidth]{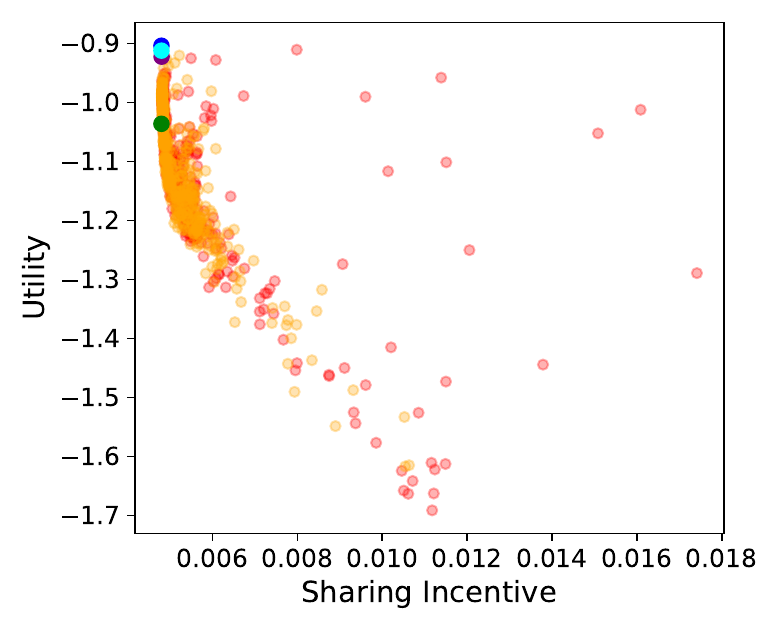}
\includegraphics[width=.32\linewidth]{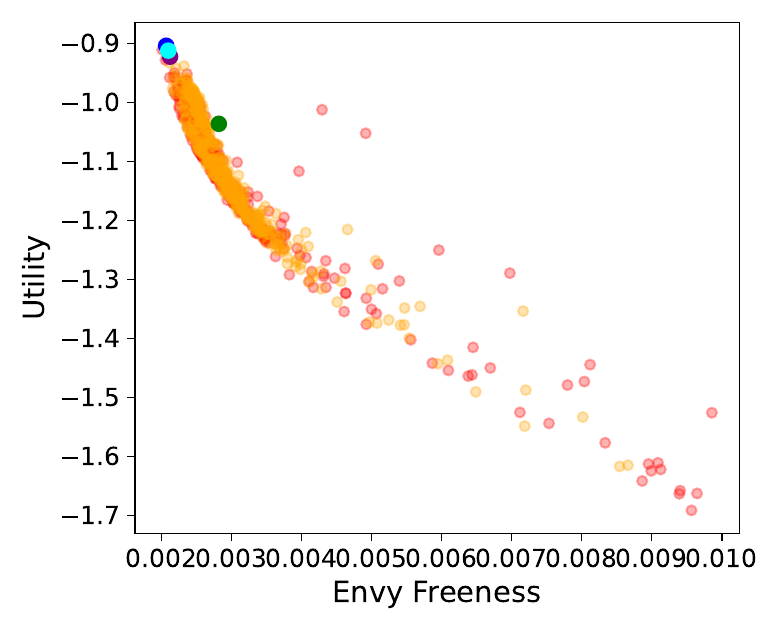}
\includegraphics[width=.32\linewidth]{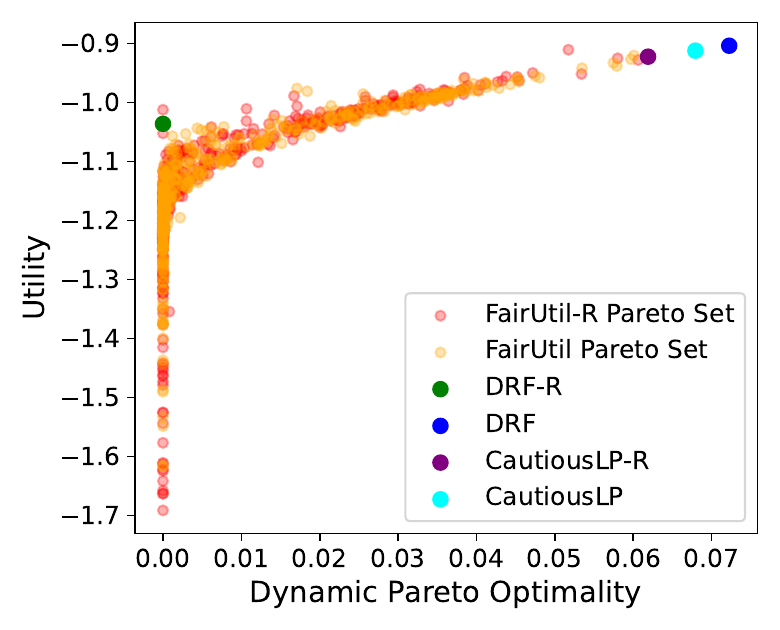}

\caption{Results on Alibaba Cluster Dataset (\texttt{openb\_pod\_list\_cpu100.csv}) with $N=20$}
\label{fig:appendix_exp_ws_alibaba_cpu}
\end{figure*}

\begin{figure*}[t]
\centering
\includegraphics[width=.32\linewidth]{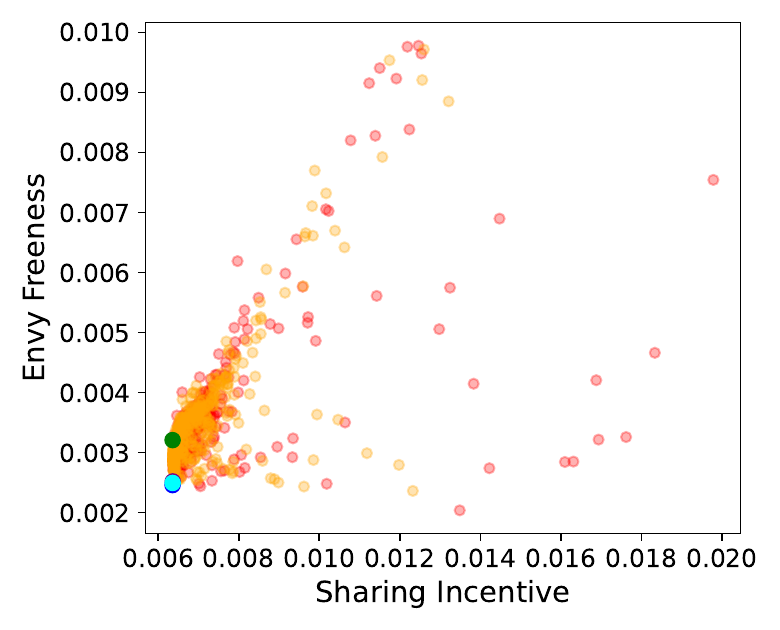}
\includegraphics[width=.32\linewidth]{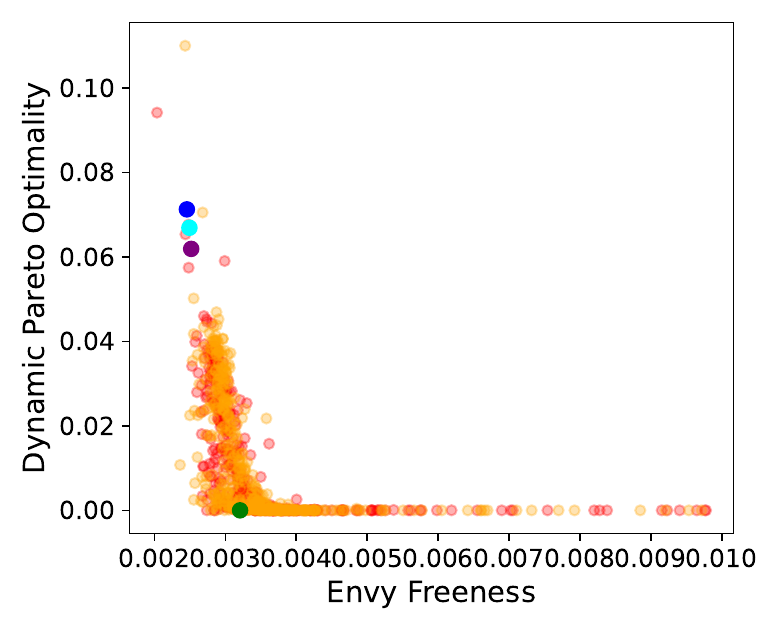}
\includegraphics[width=.32\linewidth]{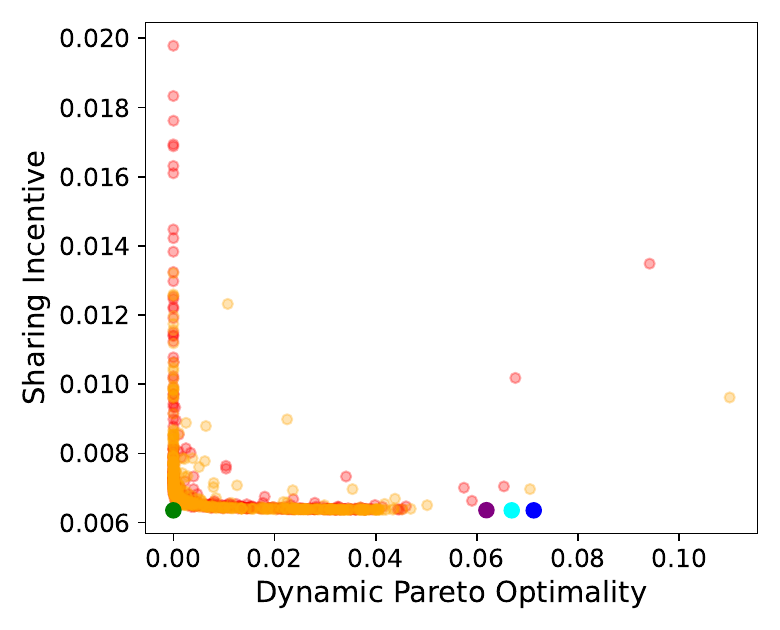}

\vspace{0.75em}

\includegraphics[width=.32\linewidth]{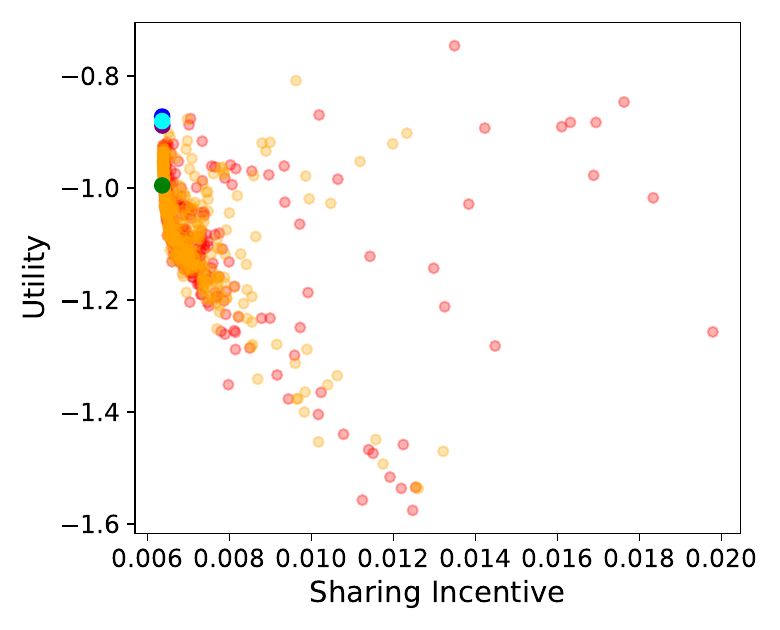}
\includegraphics[width=.32\linewidth]{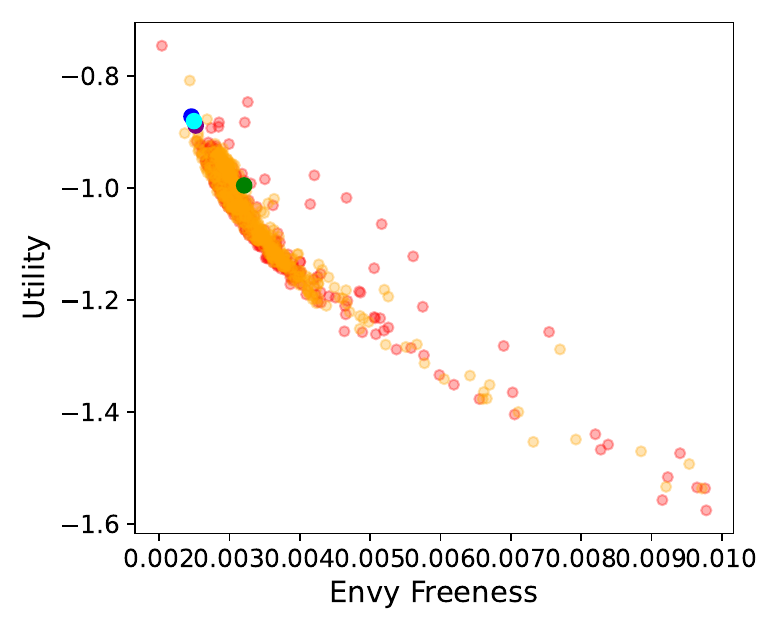}
\includegraphics[width=.32\linewidth]{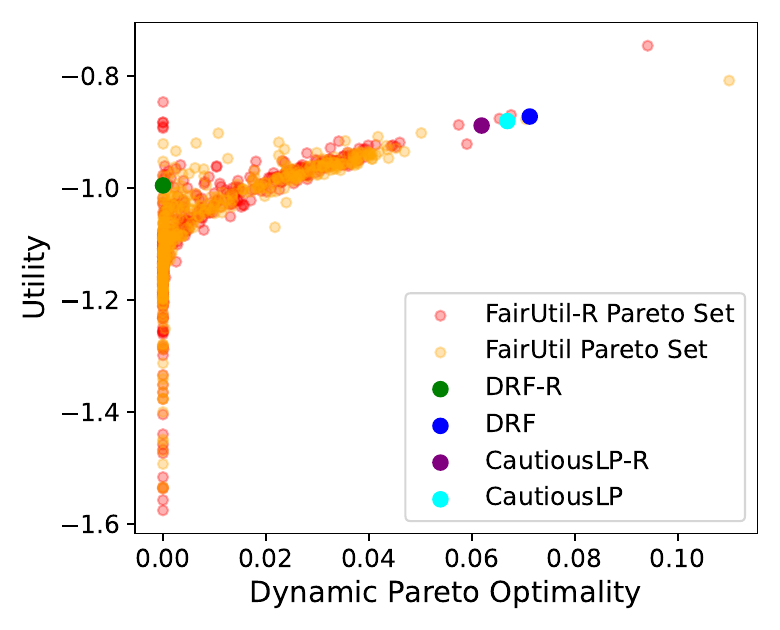}

\caption{Results on Alibaba Cluster Dataset (\texttt{openb\_pod\_list\_gpushare20.csv}) with $N=20$}
\label{fig:appendix_exp_ws_alibaba_gpu}
\end{figure*}

\begin{figure*}[t]
\centering
\includegraphics[width=.32\linewidth]{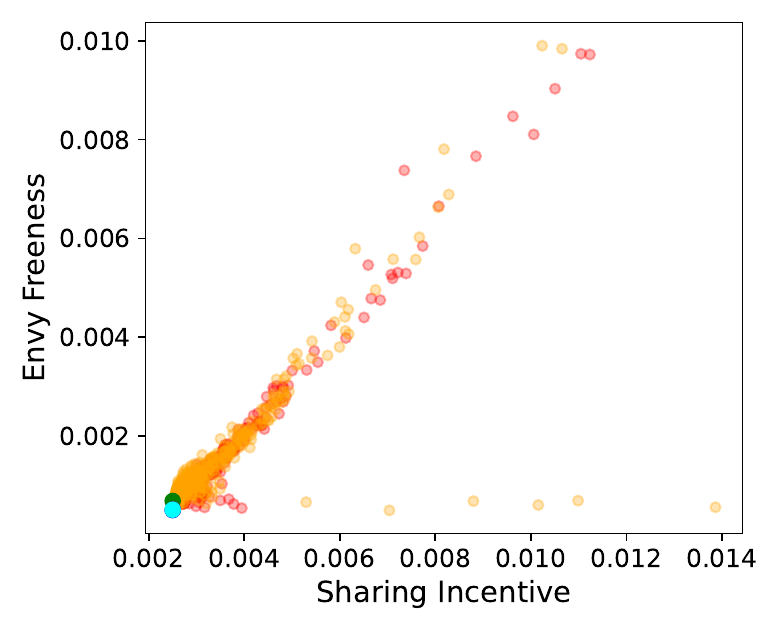}
\includegraphics[width=.32\linewidth]{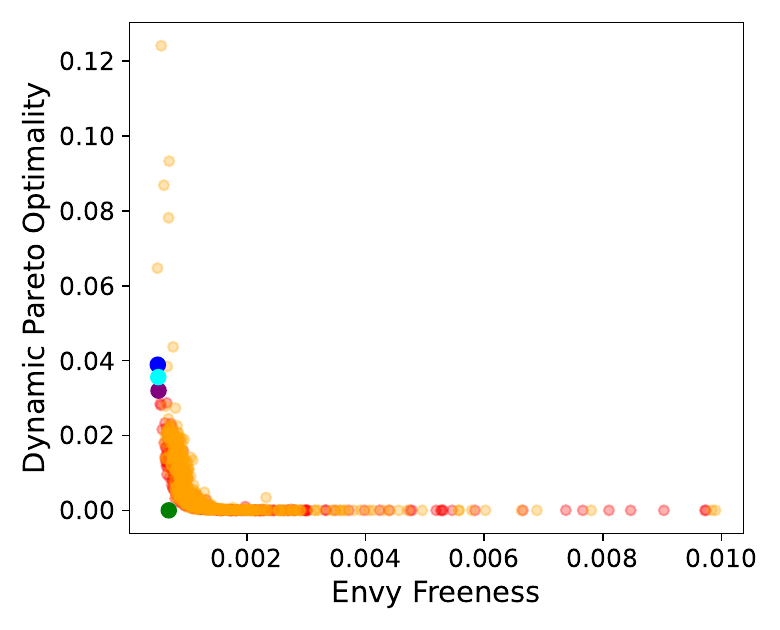}
\includegraphics[width=.32\linewidth]{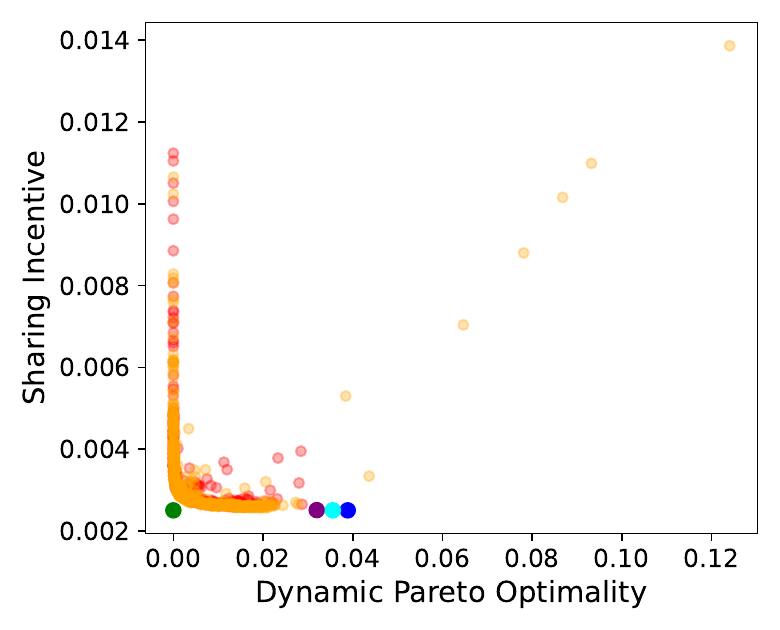}

\vspace{0.75em}

\includegraphics[width=.32\linewidth]{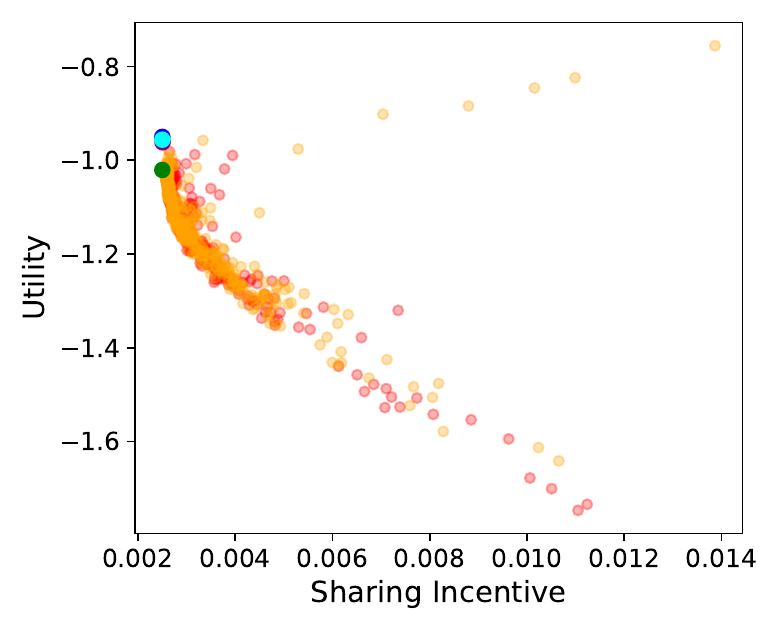}
\includegraphics[width=.32\linewidth]{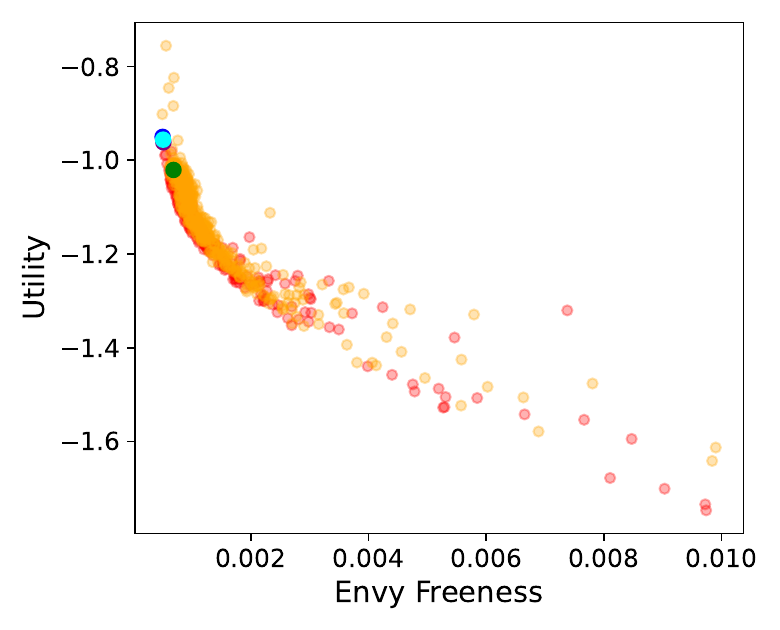}
\includegraphics[width=.32\linewidth]{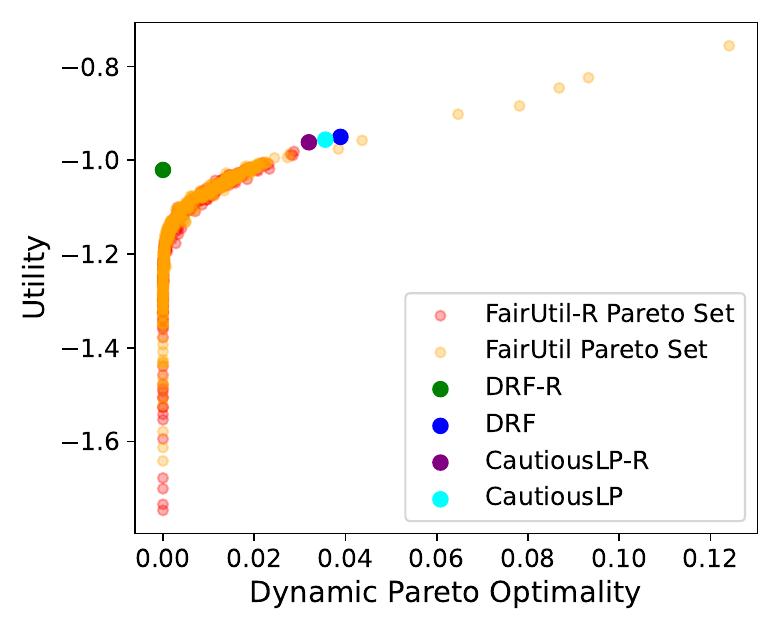}

\caption{Results on Azure dataset with $N=20$}
\label{fig:appendix_exp_ws_azure}
\end{figure*}

In the main paper, we report results with a window size of $N=10$. To further validate our approach, we extend the experiments to a larger window size of $N=20$. As shown in Figures~\ref{fig:appendix_exp_ws_alibaba_cpu}, \ref{fig:appendix_exp_ws_alibaba_gpu} and \ref{fig:appendix_exp_ws_azure}, the overall patterns remain consistent: our method continues to achieve significantly higher utility while maintaining comparable levels of fairness across all metrics. These findings demonstrate the robustness of our framework to varying window sizes and confirm its reliability under longer allocation sequences.

\begin{figure*}[t]
\centering
\includegraphics[width=.32\linewidth]{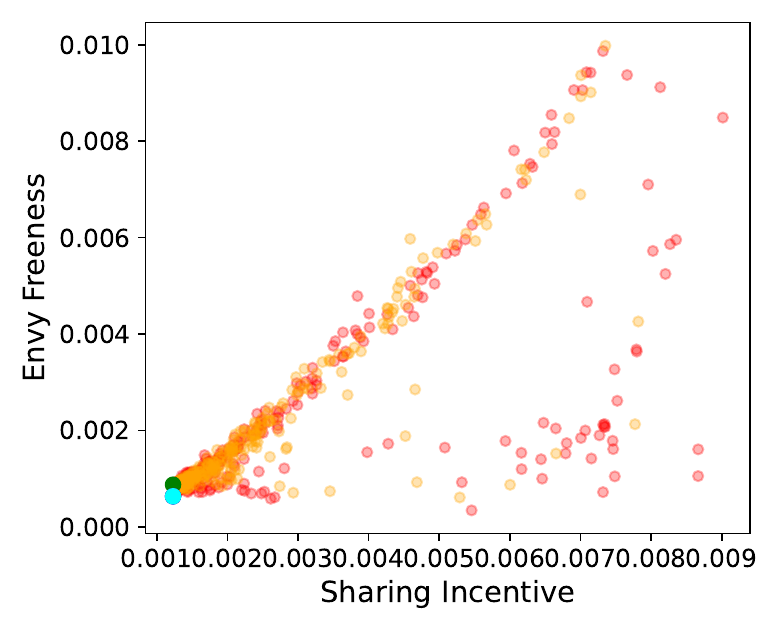}
\includegraphics[width=.32\linewidth]{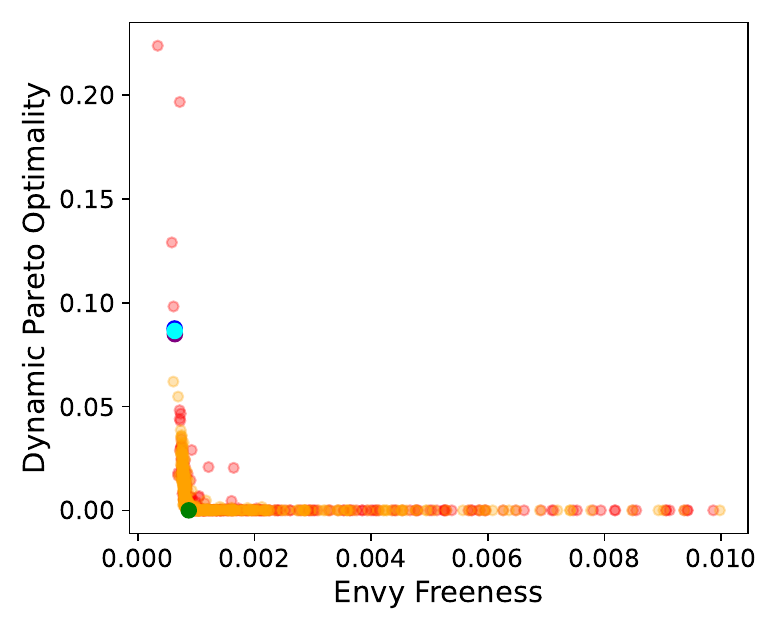}
\includegraphics[width=.32\linewidth]{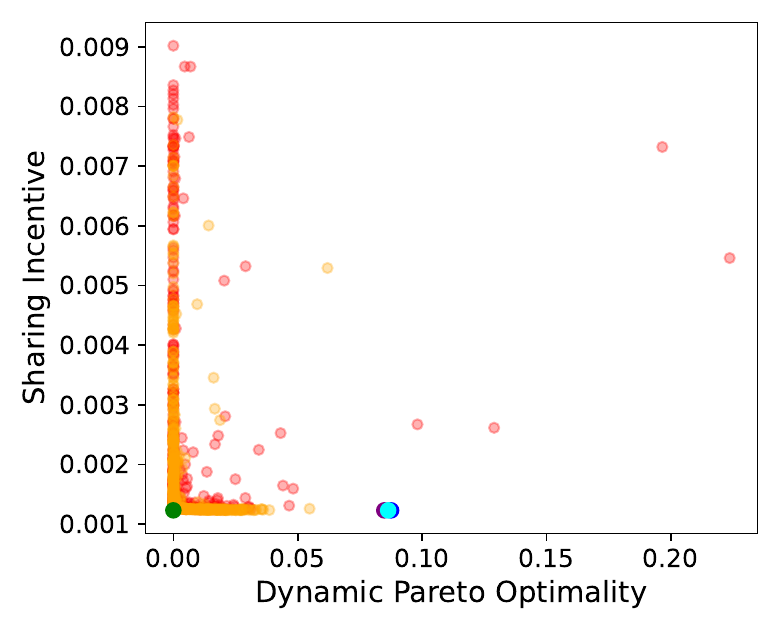}

\vspace{0.75em}

\includegraphics[width=.32\linewidth]{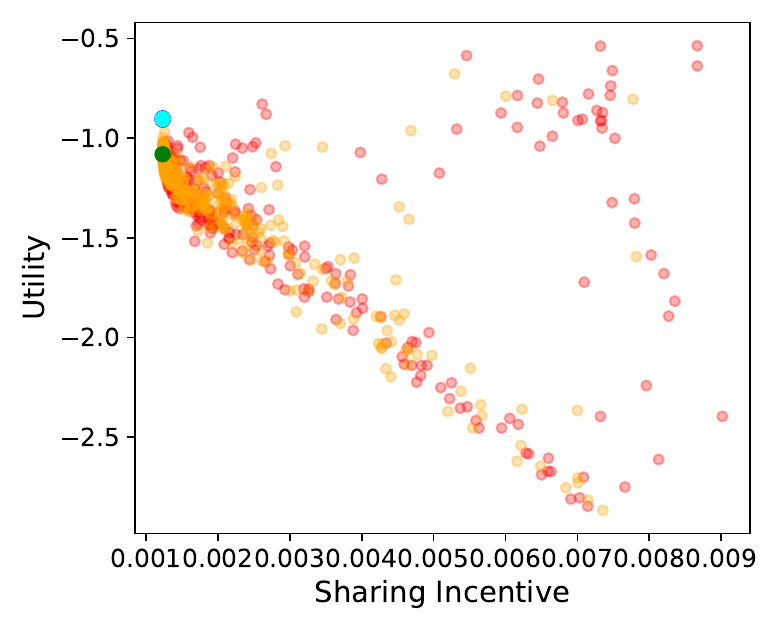}
\includegraphics[width=.32\linewidth]{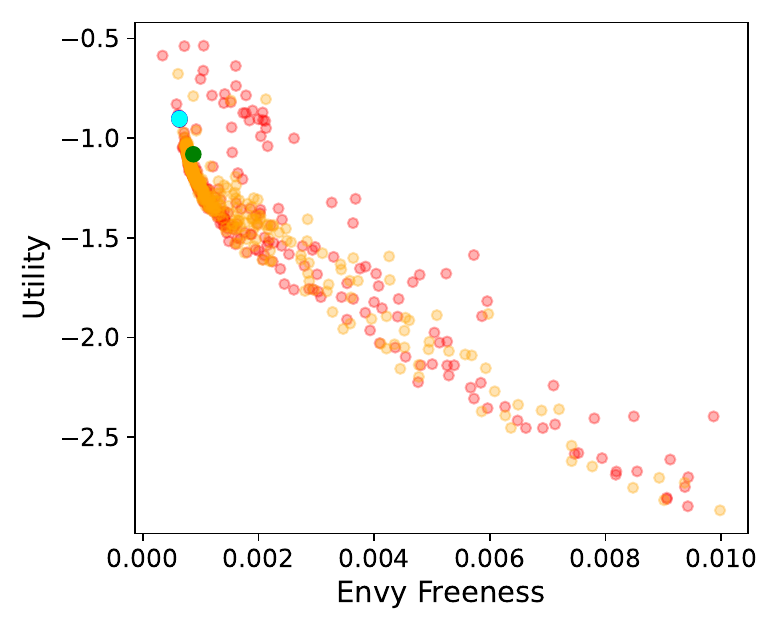}
\includegraphics[width=.32\linewidth]{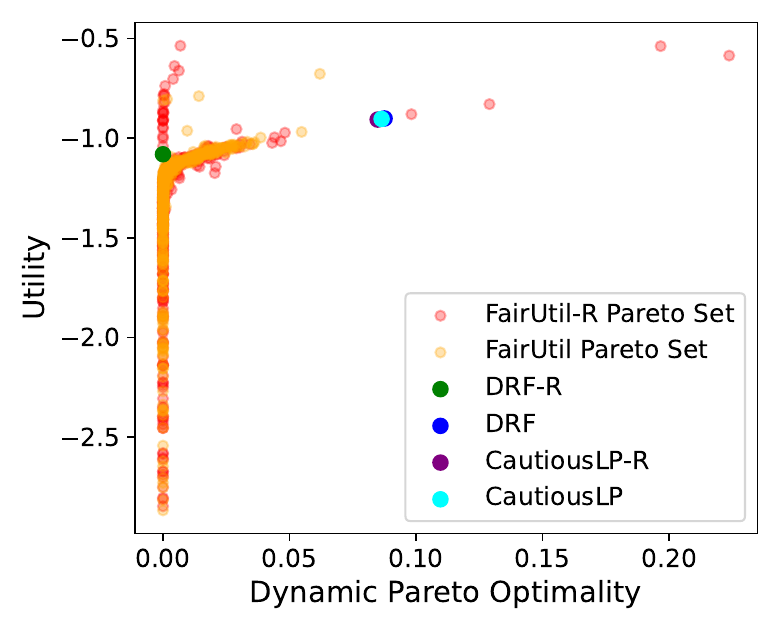}

\caption{Results on Alibaba Cluster Dataset (\texttt{openb\_pod\_list\_cpu100.csv}) with $N=80$}
\label{fig:appendix_exp_ws80_alibaba_cpu}
\end{figure*}

\begin{figure*}[t]
\centering
\includegraphics[width=.32\linewidth]{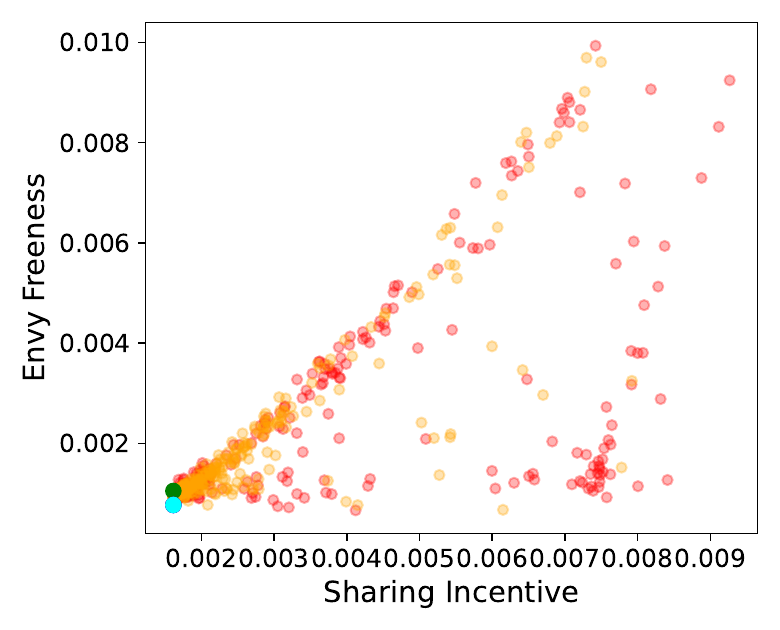}
\includegraphics[width=.32\linewidth]{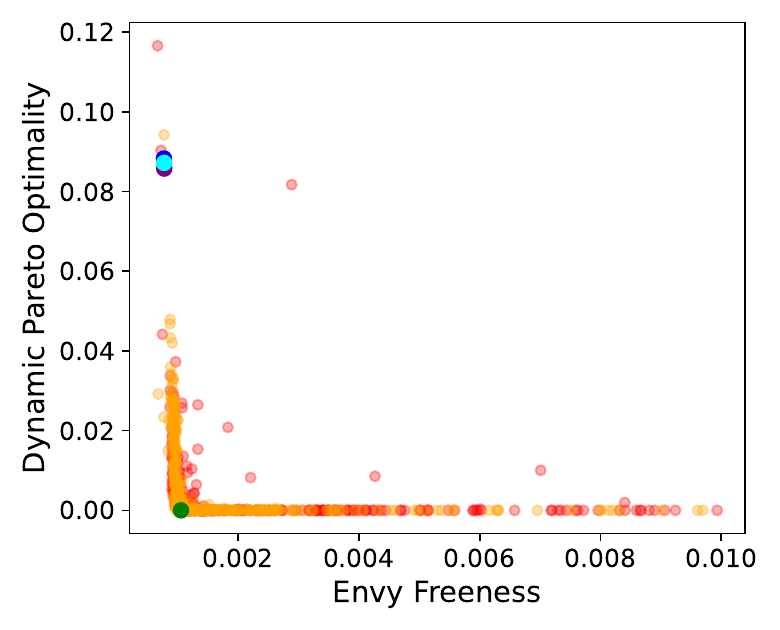}
\includegraphics[width=.32\linewidth]{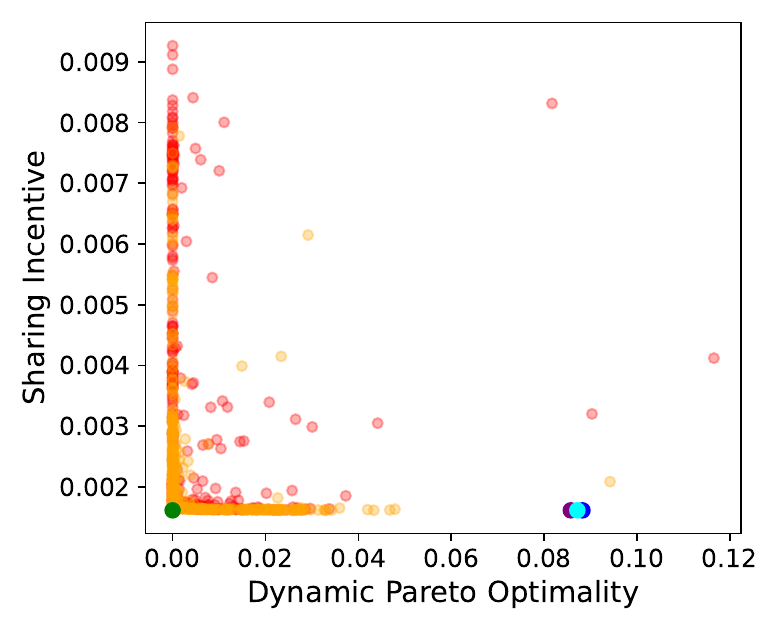}

\vspace{0.75em}

\includegraphics[width=.32\linewidth]{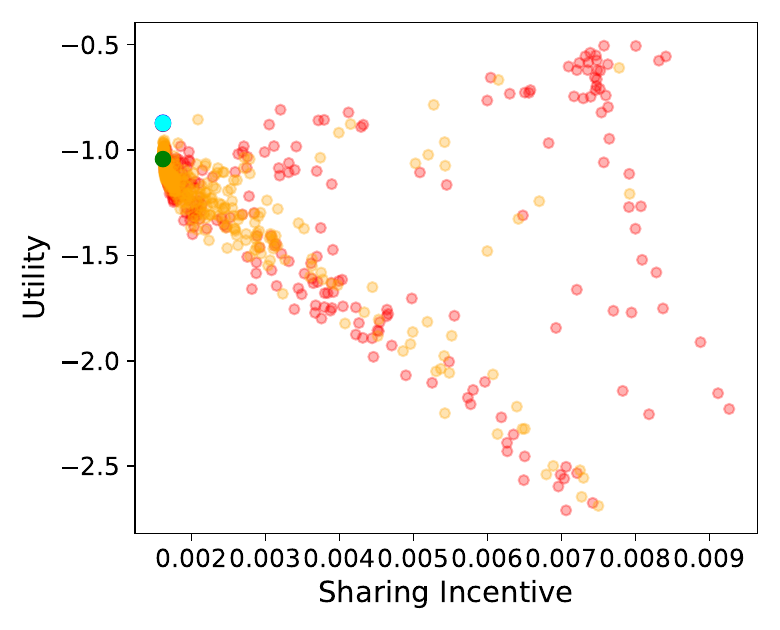}
\includegraphics[width=.32\linewidth]{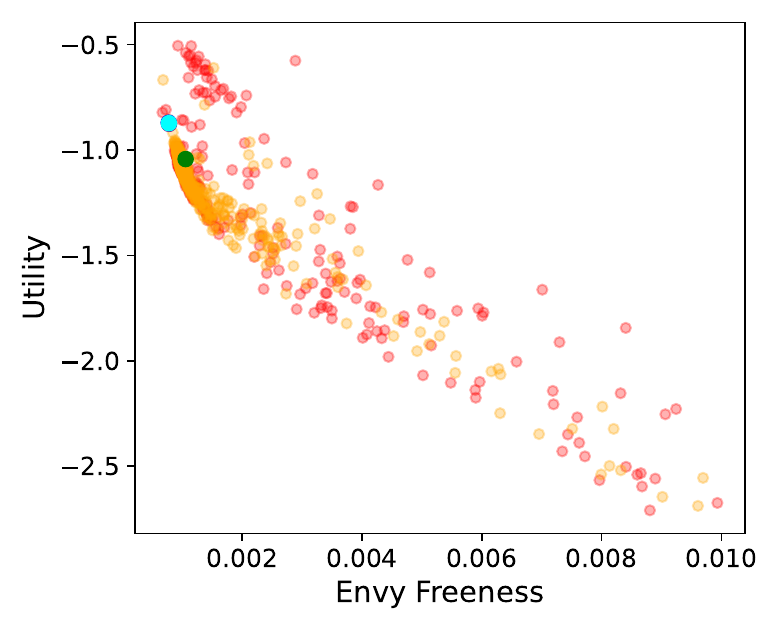}
\includegraphics[width=.32\linewidth]{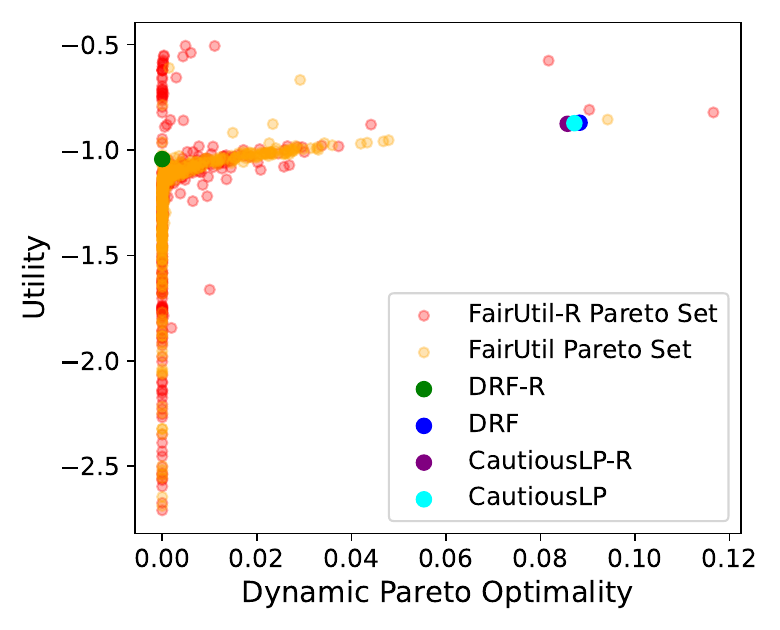}

\caption{Results on Alibaba Cluster Dataset (\texttt{openb\_pod\_list\_gpushare20.csv}) with $N=80$}
\label{fig:appendix_exp_ws80_alibaba_gpu}
\end{figure*}

\begin{figure*}[t]
\centering
\includegraphics[width=.32\linewidth]{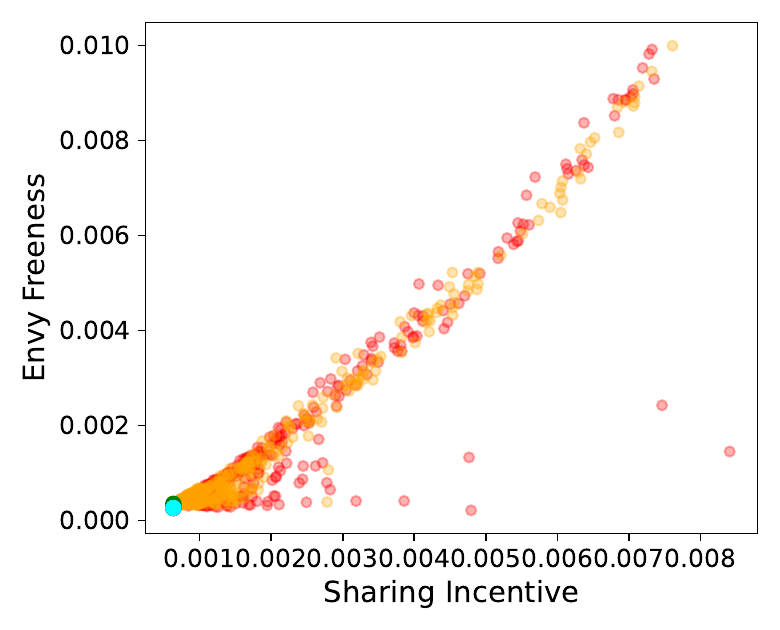}
\includegraphics[width=.32\linewidth]{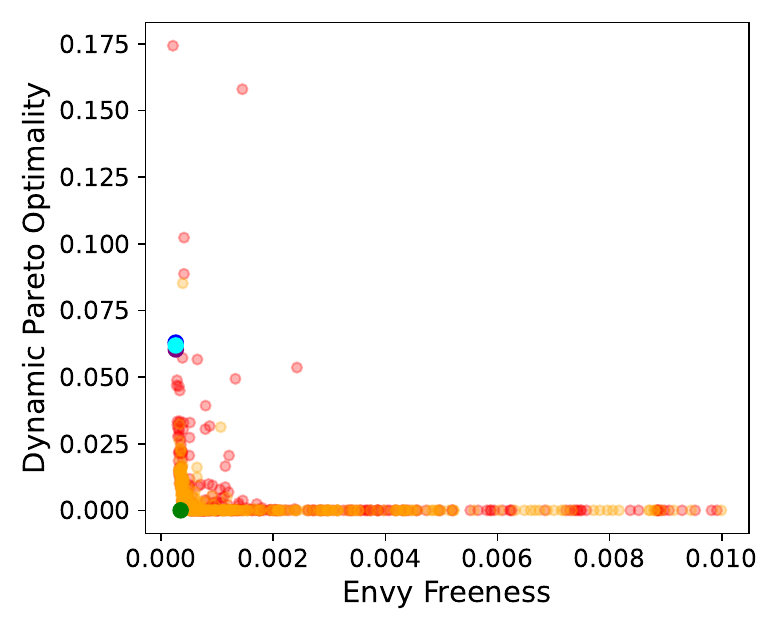}
\includegraphics[width=.32\linewidth]{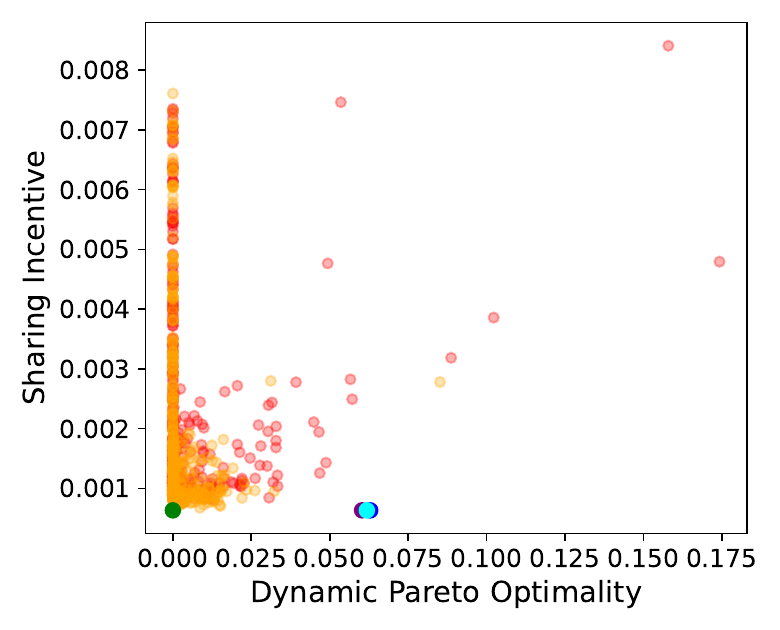}

\vspace{0.75em}

\includegraphics[width=.32\linewidth]{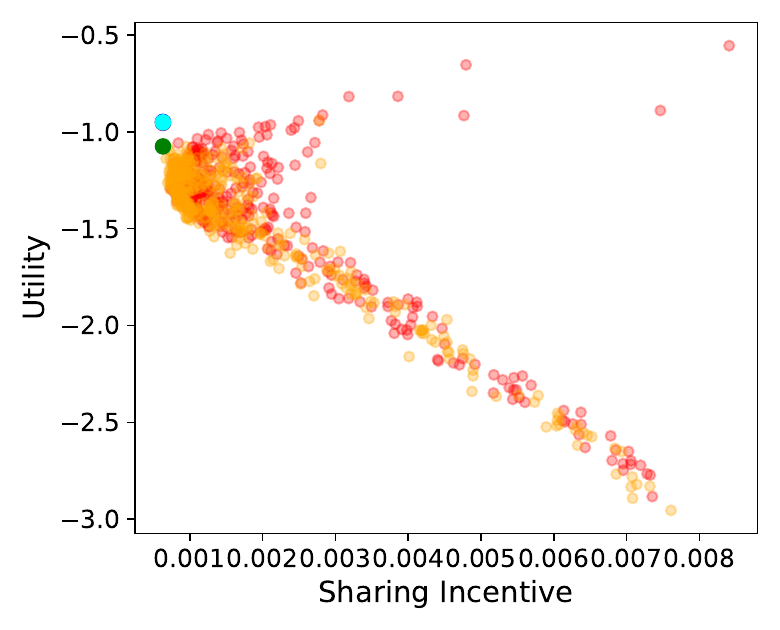}
\includegraphics[width=.32\linewidth]{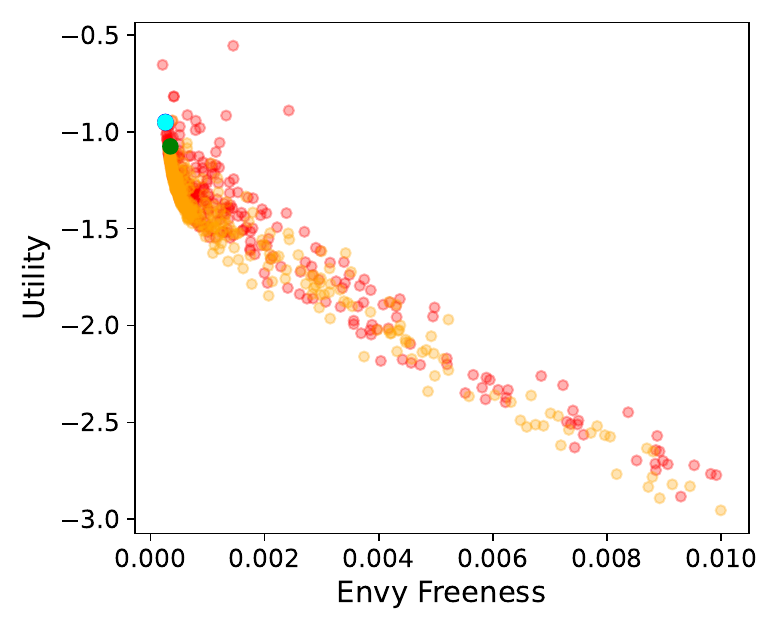}
\includegraphics[width=.32\linewidth]{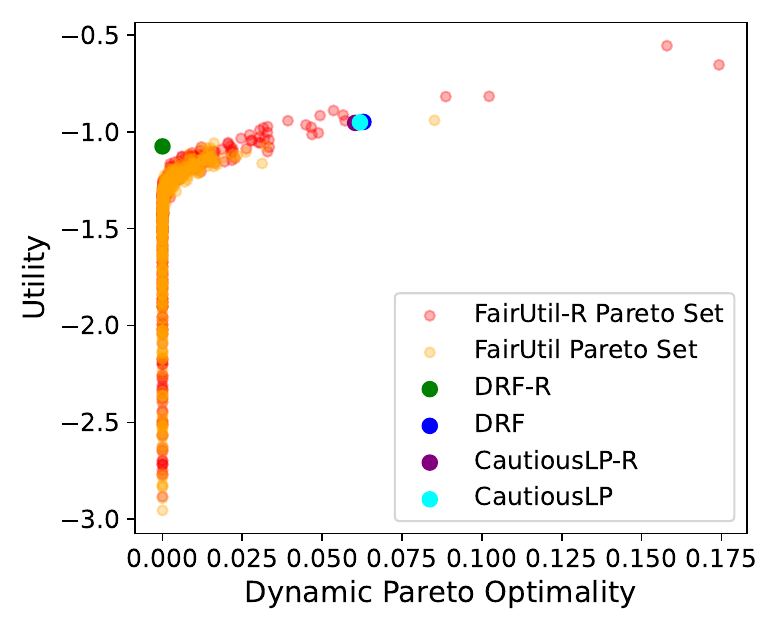}

\caption{Results on Azure dataset with $N=80$}
\label{fig:appendix_exp_ws80_azure}
\end{figure*}


\section{Source Code}
The experiment code is available at
\url{https://tinyurl.com/5xs3ukjn}




\end{document}